\documentclass[showpacs,superscriptaddress,floatfix,amsmath,amssymb,twocolumn,prb]{revtex4-1}
\usepackage{bm}
\usepackage{graphicx}
\usepackage{color}
\usepackage[normalem]{ulem}
\usepackage{hyperref}
\usepackage{enumerate}
\usepackage{amsmath}
\usepackage{epstopdf}
\usepackage{comment}
\usepackage{amssymb}

\def\sgn{{\rm sgn}}

\def\Re{{\rm Re}}

\newcommand{\addYZ}[1]{\textcolor{magenta}{#1}}

\newcommand{\DblAv}[1]{\langle\langle #1 \rangle\rangle}

\begin{document}

\title {Spectral effects of dispersive mode coupling in driven mesoscopic systems}
\author{Yaxing Zhang}
\affiliation{Department of Physics and Astronomy, Michigan State University, East Lansing, MI 48824, USA}
\author{M. I. Dykman}
\affiliation{Department of Physics and Astronomy, Michigan State University, East Lansing, MI 48824, USA}

\date{\today}

\begin{abstract}
Nanomechanical and other mesoscopic vibrational systems typically have several nonlinearly coupled modes with different frequencies and with long lifetime. We consider the power spectrum of one of these modes. Thermal fluctuations of the modes nonlinearly coupled to it lead to fluctuations of the mode frequency and thus to the broadening of its spectrum. However, the coupling-induced broadening is partly masked by the spectral broadening due to the mode decay. We show that the mode coupling can be identified and characterized using the change of the spectrum by weak resonant driving. We develop a path-integral method of averaging over the non-Gaussian frequency fluctuations from nonresonant (dispersive) mode coupling. The shape of the driving-induced power spectrum depends on the interrelation between the coupling strength and the decay rates of the modes involved. The characteristic features of the spectrum are analyzed in the limiting cases. We also find the power spectrum of a driven mode where the mode has internal nonlinearity. Unexpectedly, the power spectra induced by the intra- and inter-mode nonlinearities are qualitatively different. The analytical results are in excellent agreement with the numerical simulations.

\end{abstract}

\pacs{62.25.Fg,  85.25.-j, 78.60.Lc, 05.40.-a}

\maketitle

\section{Introduction}
\label{sec:intro}

Mesoscopic vibrational systems typically have several nonlinearly coupled modes. These can be flexural modes in the case of nanomechanical resonators\cite{Barnard2012,Eichler2012,Westra2010, Castellanos-Gomez2012,Mahboob2012a,Matheny2013,Miao2014}, photon and phonon modes in optomechanics\cite{Sankey2010, Purdy2010,Aspelmeyer2014,Singh2014,Weber2014,Paraiso2015}, or modes of microwave cavities in circuit quantum electrodynamical systems\cite{Holland2015}. Often different modes have significantly different frequencies, so that the interaction between them is primarily dispersive. A major effect of such interaction is that the frequency of one mode depends on the amplitude of the other mode. The related shift of the mode frequency provides a means of characterizing the coupling strength where both modes can be accessed, cf. Refs. \onlinecite{Westra2010,Venstra2012,Matheny2013,Vinante2014} and can be used for quantum nondemolition measurements of the oscillator Fock states.\cite{Santamore2004a,Ludwig2012} However, often only one mode can be accessed and controlled, and the presence of dispersive coupling has to be inferred from the available data.

An important consequence of dispersive coupling is that amplitude fluctuations of one mode lead to frequency fluctuations of the other mode \cite{DK_review84}. The amplitude fluctuations come from the coupling to a thermal reservoir, but they can also be of nonthermal origin. The mode-coupling induced frequency fluctuations broaden the spectrum of the response to an external force and the power spectrum. Such broadening has been suggested as a major broadening mechanism for flexural modes in carbon nanotubes \cite{Barnard2012}, graphene sheets\cite{Miao2014}, doubly clamped beams\cite{Venstra2012,Matheny2013} as well as microcantilevers.\cite{Vinante2014} 

Separating the mode-coupling induced fluctuations from other spectral broadening mechanisms is a nontrivial problem, see Ref.~\onlinecite{Sansa2015} for a recent review of the broadening mechanisms. The most familiar broadening mechanism is vibration decay due to energy dissipation.  Another mechanism of interest for the present paper is internal vibration nonlinearity. Because of such nonlinearity, the frequency of a vibrational mode depends on the mode amplitude, and thermal fluctuations of the amplitude lead to frequency fluctuations. This is reminiscent of the mode-coupling effect, yet is coming from a different type of nonlinearity. We show below how these mechanisms can be clearly distinguished.

In this paper, we propose a means for identifying and characterizing the mode-coupling induced frequency fluctuations. The  approach is based on studying the power spectrum of the considered mode in the presence of periodic driving. It does not require access to other modes and the ability to characterize them, as in Refs.~\onlinecite{Westra2010,Venstra2012,Matheny2013,Vinante2014}, for example. The approach relies on the fact that, quite generally, frequency fluctuations lead to the features in the power spectrum of a periodically driven mode, which do not occur without such fluctuations.\cite{Zhang2014}  If one thinks of the driven mode as a charged oscillator in a stationary radiation field, these features correspond to fluorescence and quasi-elastic light scattering. The absence of the latter effects in the case of a periodically driven linear oscillator with constant frequency is a textbook result.\cite{Lorentz1916,Einstein1910b,Heitler2010} 

The analysis of the power spectra of driven modes with fluctuating frequency in Ref.~\onlinecite{Zhang2014} was phenomenological. The results were obtained in some limiting cases and in the case of Gaussian fluctuations. 
The dispersive mode coupling leads to strongly non-Gaussian frequency fluctuations. The simplest type of such coupling corresponds to the coupling energy $\propto q^2q_d^2$, where $q$ is the coordinate of the considered driven mode and $q_d$ is the coordinate of the mode to which it is dispersively coupled and which we call the $d$-mode. Where the modes are far from resonance, the frequency change of the considered mode is proportional to the period-average value of $q_d^2$. Even where $q_d(t)$ is Gaussian, as in the case of thermal displacement of a linear mode,\cite{Einstein1910b} the squared displacement $q_d^2(t)$ is not. 

The goal of this paper is to develop the necessary tools and to reveal the features of the spectra related to the dispersive-coupling induced frequency fluctuations. These features depend on the interrelation between the typical magnitude of the frequency fluctuations $\Delta\omega$, their reciprocal correlation time $\Gamma_d$, and the decay rate of the considered mode $\Gamma$. We assume that all these parameters are small compared to the mode eigenfrequencies and their difference.

In the absence of driving, the power spectrum and the linear response spectrum have the form of a convolution of the spectrum of the considered mode taken separately and a function that depends on the parameter $\alpha_d = \Delta\omega/\Gamma_d$. \cite{Dykman1971} We call $\alpha_d$ the motional narrowing parameter to draw the similarity (although somewhat indirect) with the motional narrowing effect in nuclear magnetic resonance (NMR).\cite{Anderson1954,Kubo1954} For $\alpha_d\ll 1$ the correlation time of the frequency fluctuations is comparatively small. Then the fluctuations are averaged out and their effect is small, as in the case of fast decay of correlations in NMR. On the other hand, for $\alpha_d\gg 1$, the spectrum can be thought of as a superposition of partial spectra, each for a given value of the period-averaged $q_d^2$. The weight of the partial spectrum is determined by the probability density of the period-averaged $q_d^2$. 

In the presence of driving, the situation is different. The first distinction is that, without the dispersive coupling, there is no driving-induced part of the power spectrum at all, except for the trivial $\delta$-peak at the driving frequency.  Based on the previous results, \cite{Zhang2014} we expect that the driving-induced part of the spectrum will strongly depend on the interrelation between the rates $\Gamma$ and $\Gamma_d$. It is clear that it will also strongly depend on $\alpha_d$, but this dependence is not known in advance. 

The formulation below is in the classical terms. However, the results fully apply also to the case where the considered driven mode is quantum, i.e., its energy level spacing is comparable or exceed the temperature. It is essential for the analysis that the $d$-mode is classical; the case where the $d$-mode is in the deeply quantum regime is in a way simpler, as the dispersive coupling can be resolved in the power spectrum without driving.\cite{DK_review84} 

\subsection{The structure of the paper}

An important part of the paper is the averaging over the fluctuations of the $d$-mode. This is an interesting theoretical problem, with direct relevance to the experiment. It involves an explicit calculation of the appropriate path integral. The calculation is presented in Secs.~\ref{subsec:spectrum_F_dispersive} and \ref{sec:averaging}. These sections as well as Sec.~\ref{subsec:slow} can be skipped if one is interested primarily in the predictions for the experiment.  Below in Sec.~\ref{sec:driving_spectrum} we give a general expression for the power spectrum of a mode driven by an external field, which applies where the field is comparatively weak, so that the internal nonlinearity of the mode remains small. The effect of the field is most pronounced where it is close to resonance. In Sec.~\ref{sec:eom} we derive equations of motion for weakly damped dispersively coupled modes in the rotating wave approximation. Section \ref{sec:discussion} provides the explicit analytical expressions for the driving-induced part of the power spectrum in the limiting cases, which refer to the fast or slow relaxation of the $d$-mode compared to the relaxation of the driven mode and also to the large or small frequency shift due to the dispersive coupling compared to the relaxation rate of the $d$-mode. This section also presents results of numerical calculations of the spectra, which are compared with the results of simulations. The last part of the section describes the dependence of the area of the driving-induced peak on the dispersive coupling  parameters. Section \ref{sec:internal_nonlinearity} describes the driving-induced part of the power spectrum for a nonlinear oscillator, where the fluctuations of the oscillator frequency are due not to dispersive coupling but to the internal nonlinearity. The last section provides a summary of the results. The transfer-matrix method used to perform the averaging over the dispersive-coupling induced fluctuations and an outline of an alternative derivation of the major result are given in the appendices.

\section{Driving-induced part of the power spectrum}
\label{sec:driving_spectrum}

Frequency fluctuations render the mode response to driving random. Generally, the response is nonlinear in the driving strength. Respectively, even for sinusoidal driving, the mode power spectrum $\Phi(\omega)$ is a complicated function of the driving amplitude $F$ and frequency $\omega_F$. The conventional way of measuring the power spectrum of a driven system with coordinate $q$ corresponds to  the definition
\[\Phi(\omega)= 2{\rm Re}~\int_0^\infty dt e^{i\omega t}\DblAv{ q(t+t')q(t')},\] 
where $\DblAv{\cdot}$ indicates statistical averaging and averaging with respect to $t'$ over  the driving period $2\pi/\omega_F$ (sometimes the power spectrum is also defined as $\Phi(\omega)/2\pi$). If the driving is weak, one can keep in $\Phi(\omega)$ only the terms quadratic in $F$, 
\begin{equation}
\label{eq:correlator_defined}
\Phi(\omega) \approx \Phi_0(\omega) + \frac{\pi}{2}F^2|\chi(\omega_F)|^2\delta(\omega-\omega_F) +F^2\Phi_F(\omega).
\end{equation}
Here, $\Phi_0$ is the power spectrum in the absence of driving. For the considered underdamped mode, it is a resonant peak at the mode eigenfrequency $\omega_0$, which is due to thermal vibrations; the width of the peak is small compared to $\omega_0$. Function $\chi(\omega)$ is the mode susceptibility, $ \Phi_0(\omega)= (2k_BT/\omega_0){\rm Im}~\chi(\omega)$. The term $\propto |\chi(\omega_F)|^2$ describes the contribution  of stationary forced vibrations at frequency $\omega_F$.  

Of utmost interest to us is the last term in Eq.~(\ref{eq:correlator_defined}), $\Phi_F(\omega)$. For a linear oscillator with non-fluctuating frequency, this term is equal to zero. Indeed, the motion of such oscillator is a superposition of thermal vibrations and forced vibrations at frequency $\omega_F$, which are uncoupled. Frequency fluctuations  affect both thermal vibrations, leading to spectral broadening, and forced vibrations. The latter effect is particularly easy to understand in the case of slow frequency fluctuations. Here, one can think of the amplitude and phase of forced vibrations as  being determined by the  detuning of the instantaneous mode frequency from the driving frequency. Therefore they fluctuate in time, which leads to the onset of a ``pedestal" of the $\delta$-peak at $\omega=\omega_F$ in Eq.~(\ref{eq:correlator_defined}). Some other limiting cases have been considered earlier.\cite{Zhang2014}.

\section{Equations of motion for the slow variables}
\label{sec:eom}

We will consider the power spectrum $\Phi(\omega)$ in the case where the mode of interest is dispersively coupled to another mode (mode $d$). The modes are weakly coupled to a thermal reservoir, so that their decay rates are small. The mode of interest is driven by a periodic field $F\cos\omega_F t$. The Hamiltonian of the system is
\begin{align}
\label{eq:Hamiltonian}
&H=H_0+H_b+H_i, \qquad H_0=\frac{1}{2}(p^2+\omega_0^2q^2) +\frac{1}{4}\gamma q^4\nonumber\\
& +\frac{1}{2}(p_d^2+\omega_d^2q_d^2) + \frac{3}{4}\gamma_d q^2 q_d^2 -qF\cos\omega_Ft.
\end{align}
Here, $p$ and $p_d$ are the momenta of the considered mode and the $d$-mode, respectively, $\omega_0$ and $\omega_d$ are the mode eigenfrequencies, $\gamma$ is the parameter of the intrinsic nonlinearity of the considered mode, and $\gamma_d$ is the dispersive coupling parameter. We do not incorporate the intrinsic nonlinearity of the $d$-mode, as it will not affect the results if it is small, see below. 

The term $H_b$ is the Hamiltonian of the thermal bath (each mode can have its own bath, but we assume that in this case the baths have the same temperature), whereas $H_i$ describes the mode-to-bath coupling.  We assume the coupling to be linear in the modes coordinates, $H_i= qh + q_dh_d$, where $h$ and $h_d$ are functions of the dynamical variables of the bath. Such coupling is the dominant mechanism of relaxation for small mode displacements and velocities. For $H_i$ of this form, the decay rate of the considered mode is\cite{Senitzky1960,Schwinger1961}
\begin{align}
\label{eq:decay_rates}
\Gamma \equiv \Gamma(\omega_0)=\hbar^{-2}{\rm Re}\,\int_0^\infty dt\langle[h^{(0)}(t),h^{(0)}(0)]\rangle e^{i\omega_0t},
\end{align} 
where $h^{(0)}$ is function $h$ calculated in the neglect of the mode-bath coupling. The expression for the decay rate $\Gamma_d$ of the $d$-mode is similar to Eq.~(\ref{eq:decay_rates}), with $h^{(0)}$ replaced with $h^{(0)}_d$ and $\omega_0$ replaced with $\omega_d$. Parameters $\Gamma, \Gamma_d$ correspond to friction coefficients in the phenomenological description of the mode dynamics
\begin{equation}
\label{eq:eom_phenomenological}
\ddot q + \partial_qH_0 = -2\Gamma\dot q - h^{(0)}(t).
\end{equation}
The analysis below refers to slowly varying amplitudes and phases of the modes and applies even where the above phenomenological description does not apply.\cite{Senitzky1960,Schwinger1961,DK_review84}

In what follows we assume that the modes are underdamped, the nonlinearity is weak, and the driving frequency is close to resonance,
\begin{align}
\label{eq:conditions}
&\Gamma, \Gamma_d, \frac{|\gamma|\langle q^2\rangle}{\omega_0}, \frac{|\gamma_d|\langle q_d^2\rangle}{\omega_0}, |\omega_0-\omega_F| \ll
\omega_0,\omega_d, |\omega_0-\omega_d|.
\end{align}
The condition on $\gamma, \gamma_d$ means that the change of the mode frequencies due to the nonlinearity is small compared to the eigenfrequency. However, it does not mean that the effect of the nonlinearity is small, as the frequency change has to be compared with the frequency uncertainty due to the decay $\Gamma, \Gamma_d$. We assume that the nonlinear change of the $d$-mode frequency, which comes from the dispersive coupling and the internal nonlinearity of the $d$-mode, is also small compared to $\omega_d$. 

\subsection{Stochastic equations for slow variables}
\label{subsec:slow}

Where conditions (\ref{eq:conditions}) apply, the motion of the underdamped modes presents almost sinusoidal vibrations with slowly varying amplitudes and phases. It can be described by the standard method of averaging, which is similar to the rotating wave approximation in quantum optics. To this end, we change to complex variables 
\begin{align}
\label{eq:define_u}
u(t)=\frac{1}{2}[q(t)+(i\omega_F)^{-1}\dot q(t)]\exp(-i\omega_Ft)
\end{align}
and similarly $u_d(t)=\frac{1}{2}[q_d(t)+(i\omega_d)^{-1}\dot q_d(t)]\exp(-i\omega_dt)$. Disregarding fast oscillating terms in the equation for $u(t)$, we obtain
\begin{align}
\label{eq:eom}
&\dot u = -[\Gamma +i\delta\omega_F - i\xi(t)]u-i\frac{F}{4\omega_0} +f(t), 
\nonumber\\
&\delta\omega_F=\omega_F-\omega_0, \quad \xi(t)= \frac{3\gamma}{2\omega_0}|u(t)|^2 +  \frac{3\gamma_d}{2\omega_0}|u_d(t)|^2, 
\end{align}
where $f(t)=-(2i\omega_0)^{-1}h^{(0)}(t)\exp(-i\omega_Ft)$. Similarly, the equation for $u_d(t)$ reads
\begin{align}
\label{eq:u_d_eom}
\dot u_d = -\left[\Gamma_d - i\frac{3\gamma_d}{2\omega_d}|u(t)|^2)\right]u_d +f_d(t), 
\end{align}
with $f_d(t)=-(2i\omega_d)^{-1}h_d^{(0)}(t)\exp(-i\omega_dt)$.

Functions $f,f_d$ describe the forces on the modes from the thermal bath. The forces are random, and one can always choose $\langle f\rangle = \langle f_d\rangle = 0$. Asymptotically, they are delta-correlated Gaussian noises,
\begin{align}
\label{eq:noise_correlators}
&\langle f(t)f^*(t')\rangle = (\Gamma k_BT/\omega_0^2)\delta(t-t'), \nonumber\\
&\langle f_d(t)f_d^*(t')\rangle = (\Gamma_d k_BT/\omega_d^2)\delta(t-t');
\end{align}
all other correlators vanish. The $\delta$-functions here are  $\delta$-functions in ``slow" time compared to $\omega_0^{-1},\omega_d^{-1}$, and the correlation time of the thermal bath. The stochastic differential equations (\ref{eq:eom}) and (\ref{eq:u_d_eom}) are understood in the Stratonovich sense: $\langle u(t)f^*(t)\rangle = \Gamma k_BT/2\omega_0^2, \langle u_d(t)f_d^*(t)\rangle = \Gamma_d k_BT/2\omega_d^2$. These equations were obtained and their range of applicability established for a harmonic oscillator coupled to a bath,\cite{Senitzky1960,Schwinger1961} but they also hold for a weakly anharmonic oscillator.\cite{DK_review84}

As seen from the definition of the complex amplitude $u(t)$, function $\xi(t)$ in Eq.~(\ref{eq:eom}) describes a change of the frequency of the considered mode due to the intrinsic nonlinearity and the dispersive coupling. Because of the noise terms $f,f_d$, the frequency becomes a random function of time. The related frequency noise is of primary interest for this paper.

\section{The driving-induced spectrum $\Phi_F(\omega)$ for dispersive coupling}
\label{subsec:spectrum_F_dispersive}

In this and the two following sections we will study the spectrum $\Phi_F(\omega)$ in the case where the internal nonlinearity of the mode can be disregarded, i.e., one can set $\gamma=0$. In this case frequency fluctuations $\xi(t)\propto |u_d(t)|^2$ in Eq.~(\ref{eq:eom}) are due only to the dispersive nonlinear mode coupling. An important feature of this coupling is that it does not affect the frequency noise  $\xi(t)$ itself, as essentially was noticed earlier.\cite{Dykman1971} In other words, fluctuations of $|u_d(t)|^2$ are the same as if the $d$-mode were just a linear oscillator uncoupled from the considered mode. 

The simplest way to see this is to change in the equation of motion (\ref{eq:u_d_eom}) from $u_d(t)$ to $\tilde u_d(t)=K(t)u_d(t)$ with $K(t)=\exp[-3i(\gamma_d/2\omega_d)\int^t dt'|u(t')|^2 ]$. The Langevin equation for $\tilde u_d$ is $\dot{\tilde{u}}_d = -\Gamma_d\tilde u_d + \tilde f_d(t)$ with $\tilde f_d(t)=K(t)f_d(t)$. From Eq.~(\ref{eq:noise_correlators}), the noise $\tilde f_d(t)$ has the same correlation functions as $f_d(t)$. Therefore the term $\propto \gamma_d$ drops out of the equation for $\tilde u_d$. Since $|u_d|^2 = |\tilde u_d|^2$, the term $\propto \gamma_d$ does not affect $|u_d(t)|^2$ either [even though it does affect $u_d(t)$]. 

It is convenient to write $\xi(t)$ in terms of the scaled real and imaginary part of $\tilde u_d=(2\omega_0/3|\gamma_d|)^{1/2}(Q_d+ iP_d)$. Functions $Q_d,P_d$ are the scaled quadratures of a damped harmonic oscillator. They are described by the independent exponentially correlated Gaussian noises (the Ornstein-Uhlenbeck noises)\cite{Louisell1990}. Using the Langevin equation for $\tilde u_d$, we obtain 
\begin{align}
\label{eq:xi_quadratures}
&\langle Q_d(t)Q_d(0)\rangle =  \langle P_d(t)P_d(0)\rangle = \alpha_d\Gamma_d\exp(-\Gamma_d|t|),\nonumber\\ 
&\alpha_d=3|\gamma_d|k_BT/8\omega_0\omega_d^2\Gamma_d, \nonumber\\
&\xi(t)=[Q_d^2(t)+ P_d^2(t)]\sgn \gamma_d. 
\end{align}
The frequency noise of the driven mode $\xi(t)$ is non-Gaussian. Parameter $\alpha_d$ characterizes the ratio of the standard deviation of the noise, which is equal to the noise mean value $\langle \xi(t)\rangle = 3\gamma_dk_BT/4\omega_0\omega_d^2$, to its correlation rate $\Gamma_d$.

Since $\xi(t)$ is independent of $u(t)$, the Langevin equation for $u(t)$ (\ref{eq:eom}) is linear. Its solution reads
\begin{align}
\label{eq:suscept_defined}
&u(t)=\int_{-\infty}^t dt' \chi_{\rm sl}^*(t,t')[(-iF/4\omega_0) +f(t')],\nonumber\\
&\chi_{\rm sl}(t,t')=e^{-(\Gamma-i\delta\omega_F)(t-t')}\exp\left[ - i\int\nolimits_{t'}^t dt''\xi(t'')\right].
\end{align}
Function $\chi_{\rm sl}(t,t')$ describes the response of the considered driven mode to a resonant perturbation. The coefficient at $F$ averaged over realizations of $\xi(t)$ gives the resonant susceptibility of the mode,
\begin{align}
\label{eq:susceptibility}
\chi(\omega_F)=\frac{i}{2\omega_0}\int\nolimits_0^\infty dt \langle \chi_{\rm sl}(t,0)\rangle.
\end{align}
It determines the $\delta$-peak in the power spectrum (\ref{eq:correlator_defined}). From Eq.~(\ref{eq:suscept_defined}), the term $\Phi_F(\omega)$ in Eq.~(\ref{eq:correlator_defined}) for the power spectrum has the form
\begin{align}
\label{eq:Phi_F}
&\Phi_F(\omega)=(8\omega_0^2)^{-1}\Re \int\nolimits_0^{\infty} dt \exp[i(\omega-\omega_F)t]\nonumber\\
&\times \left[\int\nolimits_{-\infty}^t dt'\int\nolimits_{-\infty}^0 dt_1'\langle\chi_{\rm sl}(t,t')\chi_{\rm sl}^*(0,t_1')\rangle-4\omega_0^2|\chi(\omega_F)|^2\right]
\end{align} 
(in the integral over $t$ it is implied that Im~$\omega \to +0$).

The term $\propto f(t')$ in Eq.~(\ref{eq:suscept_defined}) determines the power spectrum of the mode $\Phi_0(\omega)$ near its maximum, $|\omega-\omega_0|\ll \omega_0$, in the absence of driving. Thus, Eqs.~(\ref{eq:suscept_defined}) - (\ref{eq:Phi_F}) give a general expression for the power spectrum of an underdamped mode with fluctuating frequency in the absence of internal nonlinearity. 

\section{Averaging over the frequency noise for dispersive coupling}
\label{sec:averaging}

To find the driving-induced part of the power spectrum $F^2\Phi_F(\omega)$, one has to perform averaging over fluctuations of $\xi(t)$ in Eq.~(\ref{eq:Phi_F}). This calculation is the central theoretical part of the paper. In what follows we outline the critical steps that are involved. 

The integrand in the expression for $\Phi_F(\omega)$ can be written as
\begin{align}
\label{eq:averaging_object}
&\langle\chi_{\rm sl}(t,t')\chi_{\rm sl}^*(0,t_1')\rangle = e^{-(\Gamma-i\delta\omega_F)(t-t') +(\Gamma+i\delta\omega_F)t_1'}\nonumber\\
&\times G^2(t,t',t_1'), \quad G^2(t, t', t_1')=\langle e^{-i\phi_\xi(t,t')+i\phi_\xi(0,t_1')}\rangle, \nonumber\\
&\phi_\xi(t,t')=\int_{t'}^tdt''\xi(t'').
\end{align}
Here, $\phi_\xi(t,t')$ is the increment of the phase of the oscillator over the time interval $(t',t)$ due to the frequency noise $\xi(t)$. Function $G$ describes the result of the averaging over the noise. 

Expression (\ref{eq:averaging_object}) can be slightly simplified using the relation (\ref{eq:xi_quadratures}) between $\xi(t)$ and the quadratures $Q_d$ and $P_d$. These quadratures are statistically independent, and therefore averaging over them can be done independently, so that
\begin{align}
\label{eq:G_function}
G(t,t',t_1')=\left\langle\exp\left[i\int_{t_0}^t dt'' k(t'')Q_d^2(t'')\right]\right\rangle, 
\end{align}
where $t_0=\min(t', t_1')$ and in the interval $t_0<t''<t$ function $k(t'')$ is equal to $0, \pm 1$,
\begin{align}
\label{eq:k_limits}
&k(t'') =
\left\{
\begin{array}{l}
 - \sgn\gamma_d, \quad \tau \leq t'' \leq t; \quad \tau = \max( t', 0) \\
 0, \quad \tau' < t'' <\tau; \quad \tau'=\max\Bigl(\min(0, t'), t_1'\Bigr)\\
\sgn(t'-t_1')\sgn\gamma_d , \quad t_0= \min(t',t_1') \leq t'' \leq \tau'.
\end{array}
\right.
\end{align}

The averaging in Eq.~(\ref{eq:G_function}) can be conveniently done using the path-integral technique. A theory of the power spectrum based on this technique was previously developed for the case where there is no driving.\cite{Dykman1971} The approach\cite{Dykman1971} can be extended to the present problem, as discussed in the Appendix~\ref{sec:app_DK_method}. However, the calculation is cumbersome. Here we use a different method, which is based on the technique \cite{Gelfand1960} for calculating determinants in path integrals.

In terms of a path integral, the mean value of a functional $L[Q_d(t)]$ of $Q_d(t)$ can be written as $\int {\cal D}Q_d(t) L[Q_d(t)]{\cal P}[Q_d(t)]$. For the considered exponentially correlated noise $Q_d(t)$, the probability density functional  is (cf. Ref.~\onlinecite{FeynmanQM}) 
\[{\cal P}[Q_d(t)]=\exp\left[-(4\alpha_d\Gamma_d^2)^{-1}\int dt(\dot Q_d + \Gamma_d Q_d)^2\right]. \] 

To find $G(t,t',t_1')$ we need to perform averaging over the values of $Q_d(t'')$ in the interval $(t_0,t)$. In a standard way, we discretize the time as $t_n=t_0+n\epsilon, n=0,\ldots,N$, where $\epsilon=(t-t_0)/N$ and $N\gg 1$. The path integral is then reduced to integrating over the values of $Q{}_n\equiv Q_d(t_n)/(4\alpha_d\Gamma_d^2\epsilon)^{1/2}$ with $1\leq n\leq N$ followed by averaging over $Q{}_0\equiv Q_d(t_0)/(4\alpha_d\Gamma_d^2\epsilon)^{1/2}$ with the Boltzmann weighting factor $\exp[-2\epsilon\Gamma_d Q{}_0^2]$.  

With the standard mid-point discretization,\cite{Phythian1977} the exponent in ${\cal P}[Q(t)]$ becomes
\begin{align}
\label{eq:discretized_exponent}
&-(4\alpha_d\Gamma_d^2)^{-1}\int dt(\dot Q + \Gamma_d Q)^2\to 
 -\sum_{n=1}^N \left[(Q{}_n -  Q{}_{n-1})^2\right.\nonumber\\ 
&\left.+\epsilon^2\Gamma_d^2  Q{}_n^2 \right] -
\epsilon\Gamma_d( Q{}_N^2 - Q{}_0^2).
\end{align}
The integral over $t''$ in Eq.~(\ref{eq:G_function}) is similarly discretized and goes over into $4\alpha_d\Gamma_d^2\epsilon^2\sum_n k_nQ{}_n^2$ with $k_n\equiv k(t_n)$. Then the expression for function $G$ becomes
\begin{widetext}
\begin{align}
\label{eq:G_averaging}
G(t,t',t_1')=I[k]/I[0], \qquad I[k]=\int dQ{}_0e^{-Q{}_0^2(1+\epsilon\Gamma_d)}\int\prod_{n=1}^N dQ{}_n\exp\left[-{\bf Q}{}^\dagger {\hat \Lambda}[k]{\bf Q}{}+ 2Q{}_0Q{}_1\right].
\end{align}
\end{widetext}
Here, ${\bf Q}{}$ is a vector with components $Q{}_1,\ldots,Q{}_N$. From Eq.~(\ref{eq:discretized_exponent}), the diagonal matrix elements of $\hat\Lambda[k]$ are  $ \Lambda_{nn}[k] = 2+ \epsilon^2 \Gamma_d^2(1 - 4i\alpha_d k_n)$ for $1 \leq n \leq N-1$,  $\Lambda_{NN}[k] = 1+ \epsilon^2 \Gamma_d^2(1-4i\alpha_d k_N)+\epsilon \Gamma_d$. The only non-zero off-diagonal matrix elements of $\hat\Lambda$ are $\Lambda_{n\,n\pm 1}[k]=-1$.

\subsection{Finding the determinant} 

The integrals in Eq.~(\ref{eq:G_averaging}) are Gaussian. Therefore the calculation of $G$ requires finding the determinants of the matrices $\hat \Lambda[k]$ and $\hat\Lambda[0]$. This can be done following the approach \cite{Gelfand1960}.  We consider the determinant $D_n\equiv D_n[k]$ of the square submatrix of $\hat\Lambda[k]$, which is located at the lower right corner and has rank $N-n+1$. For example, $D_1$ is the determinant of the whole matrix $\hat \Lambda$, whereas $D_N$ is the matrix element $\Lambda _{NN}$. The result of integration over $Q{}_1,\ldots,Q{}_N$ in Eq.~(\ref{eq:G_averaging}) for $I[k]$, besides the $Q_0$-dependent factor discussed below, is $\pi^{N/2}/\sqrt{D_1[k]}$.

It is straightforward to see that $D_n$ satisfies the recurrence relation
\begin{equation}
\label{eq:recurrence}
D_n = [2+\epsilon^2 \Gamma_d^2(1-4i\alpha_dk_n)] D_{n+1} - D_{n+2}, \quad 1 \leq n \leq N-2.
\end{equation}
In the limit $\epsilon \to 0$, $D_n[k]$ goes over into $D(t_n;k)$ and Eq.~(\ref{eq:recurrence}) reduces to a differential equation for $D(t'')\equiv D(t''; k)$,
\begin{equation}
\label{eq:ddot_determinant}
\ddot D(t'') - \Gamma_d^2[1-4i\alpha_dk(t'')] D(t'') = 0
\end{equation}
The obvious boundary conditions for function $D$ are $D(t) = \lim_{\epsilon \rightarrow 0} D_N = 1$ and $\dot D(t) = \lim_{\epsilon \rightarrow 0} (D_N - D_{N-1})/\epsilon = -\Gamma_d$. The quantity of interest is $D_1[k]\approx D(t_0;k)$; we will also need $\dot D(t_0;k)$, see below.

The integration of the linear in $Q{}_1$ term in the exponent in Eq.~(\ref{eq:G_averaging}) gives the factor  $\exp[Q{}_0^2 (\hat\Lambda^{-1})_{11}]$. It follows from the above analysis that $ (\hat\Lambda^{-1})_{11} = D(t_0+2\epsilon)/D(t_0+\epsilon)\approx 1+\epsilon \dot D(t_0)/D(t_0)$. Then the result of integration over $Q{}_0$ in Eq.~(\ref{eq:G_averaging})  is the factor $\{\pi/\epsilon [\Gamma_d - \dot D(t_0; k)/D(t_0;k)]\}^{1/2}$ in $I[k]$. 

For $k=0$, we have from Eq.~(\ref{eq:ddot_determinant}) $D(t'';0)=  \exp[\Gamma_d(t-t'')]$. With this, the expression (\ref{eq:G_averaging}) for function $G$ becomes
\begin{align}
\label{eq:G_final}
G(t,t',t_1')= \left\{2\Gamma_de^{\Gamma_d(t-t_0)}/[\Gamma_d D(t_0;k)-\dot D(t_0;k)]\right\}^{1/2}
\end{align}

This expression is the central result of the section. It reduces the problem of calculating the driving-induced part of the power spectrum to solving an ordinary differential equation (\ref{eq:ddot_determinant}).

\subsection{The average susceptibility}

We start the discussion of the applications of the general result (\ref{eq:G_final}) with the analysis of the factor  $\langle \chi_{\rm sl}(t,0)\rangle$, which gives the average mode susceptibility, Eq.~(\ref{eq:susceptibility}). Function  $\langle \chi_{\rm sl}(t,0)\rangle$ is given by $\langle \exp[-i\int_0^t dt''\xi(t'')]\rangle$, which in turn is given by Eq.~(\ref{eq:averaging_object}) with $G$ of the form of  Eq.~(\ref{eq:G_function}) in which $t'=t_1'=0$ and $k(t'')=-\sgn\gamma_d$. Solving Eq.~(\ref{eq:ddot_determinant}) with $k(t'')=$~const is straightforward, as is also finding then $G(t,0,0)$ from Eq.~(\ref{eq:G_final}). The result reads
\begin{align}
\label{eq:chi_explicit}
&\langle \chi_{\rm sl}(t,0)\rangle =\exp\left[-(\Gamma-i\delta\omega_F)t\right] \tilde\chi(t),\qquad \tilde\chi(t)=e^{\Gamma_d t} \nonumber\\
& \times\left[\cosh a_d t + (\Gamma_d/a_d)(1+2i\alpha_d\sgn\gamma_d)\sinh a_dt\right]^{-1}, \nonumber\\
&a_d = \Gamma_d(1+4i\alpha_d\sgn\gamma_d)^{1/2}.
\end{align} 
Equation (\ref{eq:chi_explicit}) expresses the average susceptibility in elementary functions. It agrees with the result\cite{Dykman1971} for the correlation function of the mode dispersively coupled to a fluctuating mode.

\subsection{The average of the susceptibilities product}
\label{sec:app_G_explicit}

Solving the full Eq.~(\ref{eq:ddot_determinant}) with discontinuous $k(t'')$ is more complicated. It can be done by finding function $D(t'')$ piecewise where $k(t'')=$~const as a sum of ''incident" and ''reflected" waves and then matching the solutions. The corresponding method reminds the transfer matrix method. It is described in Appendix \ref{sec:transfer_matrix}. The result reads
\begin{align}
\label{eq:result_linear_oscillator}
&[G(t,t',t_1')]^{-2} = [\tilde\chi_1(\tau_1)\tilde\chi_3(\tau_3)]^{-1} - \frac{(\Gamma_d^2-a_1^2)(\Gamma_d^2-a_3^2)}{4\Gamma_d^2a_1a_3} \nonumber\\ 
&\times \sinh (a_1\tau_1) \,\sinh (a_3\tau_3) \,\exp[-\Gamma_d(2\tau_2+\tau_1+\tau_3)]
\end{align}
where $\tau_1 = t-\tau, \tau_2 = \tau-\tau', \tau_3 = \tau' - t_0$; parameters $t_0$, $\tau$, and $\tau'$ are expressed in terms of $t,t', t_1'$ in Eq.~(\ref{eq:k_limits}). In Eq.~(\ref{eq:result_linear_oscillator}),  $\tilde\chi_1(t) = \tilde\chi(t)$, where $\tilde\chi(t)$ is given by Eq.~(\ref{eq:chi_explicit}); function $\tilde\chi_3(t)=  \tilde\chi(t)$  and $a_3 = a_1\equiv a_d$ for $t'< t_1'$, whereas   $\tilde\chi_3(t)=\tilde\chi^*(t)$ and $a_3=a_d^*$ for $t'>t_1'$.

This expression is unexpectedly simple. We use it below for analytical calculations, in particular  for calculating the driving-induced part of the power spectrum in the limiting cases.

\section{Discussion of  results}
\label{sec:discussion}

In this section we use the above results to discuss the form of the driving-induced part $F^2\Phi_F(\omega)$ of the power spectrum in the case of dispersive coupling. We give explicit expressions for the spectrum in the limiting cases. We also present the results of the numerical calculations of the spectrum based on the general expressions (\ref{eq:Phi_F}), (\ref{eq:averaging_object}), and (\ref{eq:result_linear_oscillator}). These results are compared with the simulations.
The simulations were performed in a standard way by integrating the stochastic equations of motion (\ref{eq:eom}) - (\ref{eq:noise_correlators}) using the Heun scheme \cite{Mannella2002a}.

 As we show, the shape and the magnitude of $\Phi_F(\omega)$ sensitively depend on two factors. One is the interrelation between the magnitude of frequency fluctuations $\Delta\omega$ and their bandwidth $2\Gamma_d$, the motional-narrowing parameter $\alpha_d$ defined in Eq.~(\ref{eq:xi_quadratures}). The other is the interrelation between $\Gamma_d$ and the width of the mode spectrum in the absence of driving $\Phi_0(\omega)$.  The latter is not given by just the mode decay rate $\Gamma$. It is affected by the frequency noise and depends on $\alpha_d$. 

Indeed, the width of the mode spectrum is determined by the susceptibility Im~$\chi(\omega)$. The peak of  Im~$\chi(\omega)$ is described by Eqs.~(\ref{eq:susceptibility}) and (\ref{eq:chi_explicit}), cf. also Ref.~\onlinecite{Dykman1971}. The halfwidth of this peak varies with the frequency noise strength $\alpha_d\Gamma_d$ from $\Gamma + 2\alpha_d^2\Gamma_d$, for $\alpha_d\ll 1$, to $\sim \Gamma + 2\alpha_d\Gamma_d$ for $\alpha_d\gg 1$ (the peak of Im~$\chi(\omega)$ is profoundly non-Lorentzian for $\alpha_d\gg 1$ and $\alpha_d\Gamma_d\gtrsim \Gamma$). The spectrum $\Phi_F(\omega)$ allows one to measure both the strength and the correlation time of the frequency noise and, in the first place, to directly identify the very presence of this noise.

\subsection{The spectrum $\Phi_F(\omega)$ in the limiting cases}

\subsubsection{Weak frequency noise}

The general expression for the spectrum simplifies in the limiting cases where the frequency noise is weak or its bandwidth is large or small compared to the width of the driven mode spectrum $\Phi_0 (\omega)$. In the case of dispersive coupling, the limit of weak frequency noise is realized where $\alpha_d\Gamma_d\ll \Gamma$. A general expression for $\Phi_F(\omega)$ for weak frequency noise was obtained earlier \cite{Zhang2014}. It relates $\Phi_F(\omega)$ to the power spectrum of the frequency noise $\xi(t)$. From Eq.~(\ref{eq:xi_quadratures}), in the present case we have $\langle \xi(t)\rangle = 2\langle Q_d^2(t)\rangle = 2\alpha_d\Gamma_d$, whereas the correlation function of the noise increment  $\delta \xi(t)=\xi(t)-\langle \xi(t)\rangle$ is $4\alpha_d^2\Gamma_d^2\exp(-2\Gamma_d|t|)$. Then, extending the results \cite{Zhang2014} to noise with non-zero average, we obtain
\begin{align}
\label{eq:weak_noise}
&\Phi_F(\omega)= (\alpha_d^2\Gamma_d^3/\omega_0^2)[(\omega_F-\tilde\omega_0)^2 + \Gamma^2]^{-1}
\nonumber\\
&\times \{[(\omega-\omega_F)^2+ 4\Gamma_d^2] [(\omega-\tilde\omega_0)^2 + \Gamma^2]\}^{-1},
\end{align}
where $\tilde\omega_0=\omega_0+\langle\xi(t)\rangle=\omega_0+ 2\alpha_d\Gamma_d$.

From Eq.~(\ref{eq:weak_noise}), for weak dispersive-coupling induced noise, the intensity of the spectrum $\Phi_F(\omega)$ is proportional to the square of the coupling parameter. If the detuning of the driving field frequency from the eigenfrequency of the driven oscillator largely exceeds the half widths of the power spectra of the both oscillators in the absence of driving, $|\omega_F-\tilde \omega_0|\gg \Gamma,\Gamma_d$, the power spectrum $\Phi_F(\omega)$ has two distinct peaks. One is located at the  oscillator frequency $\tilde\omega_0$ and has halfwidth $\Gamma$. The other is located at the driving frequency $\omega_F$ and has halfwidth  $2\Gamma_d$. We note that the noise  variance $4\alpha_d^2\Gamma_d^2$ is independent of $\Gamma_d$. For constant $\alpha_d^2\Gamma_d^2$, the areas of the peaks at $\tilde \omega_0$ and $\omega_F$ are $\propto \Gamma_d/\Gamma\delta\omega_F^4$ and $\delta\omega_F^{-4}$, respectively (the ratio $\Gamma_d/\Gamma$ affects only the area of the peak at $\omega_0$). As $|\delta\omega_F|$ decreases, the peaks start overlapping and for small $|\delta\omega_F|$ may not be resolved. This behavior is general and occurs also where the frequency noise is not weak, as we show below.

\subsubsection{Broad-band frequency noise}

We now consider the case of the broad-band frequency noise, where the decay rate of the $d$-mode $\Gamma_d$ largely exceeds the width of the driven mode spectrum.  This condition requires that the motional-narrowing parameter be small, $\alpha_d\ll 1$. At the same time, the contribution of the frequency noise to the spectrum width of the driven mode does not have to be small compared to the decay rate of this mode, i.e., the ratio $\alpha_d^2\Gamma_d/\Gamma$ can be arbitrary. In other words, the broadening of the spectrum of the driven mode can still largely come from the dispersive coupling.

For large $\Gamma_d$ and small $\alpha_d$, Eq.~(\ref{eq:ddot_determinant}) can be solved in the WKB approximation, $D(t'')\approx \exp\{-\Gamma_d\int_t^{t''} dt_2 [1-4i\alpha_dk(t_2)]^{1/2}\}$. Combined with Eq.~(\ref{eq:G_final}), this solution immediately gives the averaging factor $G$ in Eq.~(\ref{eq:averaging_object}) and thus the spectrum $\Phi_F$. Alternatively, one can use the explicit expression (\ref{eq:result_linear_oscillator}) for function $G$. Only the first term has to be kept in this expression for $\alpha_d\ll 1$. Integration over time in Eq.~(\ref{eq:Phi_F}) gives 
\begin{align}
\label{eq:broad_band}
&\Phi_F(\omega)=\frac{2\alpha_d^2\Gamma_d/\Gamma}{4\omega_0[\tilde\Gamma ^2 +(\omega_F -\tilde\omega_0)^2]}
{\rm Im}~\chi(\omega), \nonumber\\
&\chi(\omega)=(i/2\omega_0)[\tilde\Gamma -i(\omega-\tilde\omega_0)]^{-1}. 
\end{align}
Parameters $\tilde\Gamma$ and $\tilde\omega_0$ are the halfwidth of the spectrum and the eigenfrequency of the driven mode renormalized due to the dispersive coupling,  $\tilde\Gamma = \Gamma + 2\alpha_d^2\Gamma_d$.

Equation (\ref{eq:broad_band}) shows that, for a broad-band frequency noise, the spectrum $\Phi_F(\omega)$ has the same shape as the spectrum in the absence of driving, $\Phi_F(\omega)\propto \Phi_0(\omega) \propto {\rm Im}~\chi(\omega)$. However, from the spectrum $\Phi_0(\omega)$, which is Lorentzian in this case, one cannot tell whether the spectrum halfwidth $\tilde\Gamma$ is due to decay or to frequency fluctuations. In contrast, $\Phi_F\propto\alpha_d^2$ is proportional to the variance of the frequency noise and has a characteristic temperature dependence ($\alpha_d^2\propto T^2$ if $\Gamma_d$ is $T$-independent). It enables identifying the frequency noise contribution to the spectral broadening, as it was earlier demonstrated for a $\delta$-correlated frequency noise.\cite{Zhang2014}\addYZ{\sout.}  We emphasize that the full driving-induced term in the power spectrum $F^2\Phi_F(\omega)$ can be seen even where thermal fluctuations of the driven mode are weak and the peak in the power spectrum $\Phi_0(\omega)$ is too small to be resolved.

\subsubsection{Narrow-band frequency noise}

The spectrum $\Phi_F(\omega)$ has a characteristic shape also in the opposite limit where the bandwidth of the frequency noise $2\Gamma_d$ is small compared to the width of the spectrum in the absence of driving $\Phi_0(\omega)$. In this case $\Phi_F(\omega)$ displays a characteristic peak at the driving frequency $\omega_F$, as can be already inferred from the weak-noise expression (\ref{eq:weak_noise}). In the overall spectrum $\Phi(\omega)$ it looks like a pedestal of the $\delta$-peak at $\omega_F$. The typical halfwidth of the pedestal is given by $\Gamma_d$. This allows one to read off the decay rate of the $d$-mode from the spectrum $\Phi_F(\omega)$ without accessing the $d$-mode directly. To resolve the pedestal for large $\alpha_d$, where $\alpha_d\Gamma_d \gtrsim \Gamma$ so that the width of the spectrum $\Phi_0(\omega)$ is primarily determined by the dispersive coupling, we need a comparatively large detuning of the driving frequency from the maximum of $\Phi_0(\omega)$.

The simplest way to find $\Phi_F(\omega)$ near $\omega_F$ for small $\Gamma_d$ is based on Eq.~(\ref{eq:result_linear_oscillator}).  One first notices from Eq.~(\ref{eq:Phi_F}) that the major contribution to $\Phi_F(\omega)$ in this case comes from the time range $t\sim \Gamma_d^{-1}$, whereas  $t-t',|t_1'|\lesssim |\Gamma  -i\delta\omega_F|^{-1}$. Therefore in Eq.~(\ref{eq:result_linear_oscillator}) one is interested in the limit of large $t$ but comparatively small $t-t',|t_1'|$.  In the limit $t\to \infty$ but for fixed $t-t'$, function $G^{-2}(t,t',t_1')\to 1/\tilde\chi(t-t')\tilde\chi^*(-t_1')$ with $\tilde\chi(t)$ given by Eq.~(\ref{eq:chi_explicit}). The remaining term in $G^{-2}$ is $\propto \exp[-\Gamma_d (t+t')]$. One can then write the integrand in the expression (\ref{eq:Phi_F}) for $\Phi_F(\omega)$ as a series in $\exp[-\Gamma_d(t+t')]$. The next simplification comes from the fact  that $|a_d (t-t')|, |a_dt_1'|\ll 1$ for  $|\Gamma- i\delta\omega_F|/|a_d|\gg 1$. Therefore one can expand $\tilde\chi(t-t')\approx [1+2i\alpha_d\Gamma_d(t-t')]^{-1}$ and similarly for $\tilde\chi^*(-t_1')$. Ultimately, the result of integration over the times $t,t', t_1'$ reads
\begin{align}
\label{eq:slow_noise_simplify}
&\Phi_F(\omega) = \sum_{n=1}^\infty \left|\frac{1}{n!}\alpha_d^n\frac{\partial^n}{\partial\alpha_d^n}\chi(\omega_F)\right|^2 \frac{n\Gamma_d}{(\omega-\omega_F)^2+ (2n\Gamma_d)^2}. 
\end{align}
This expression describes the spectral peak at small $|\omega-\omega_F|$ in terms of the derivatives of the susceptibility $\chi(\omega)$ calculated as a function of the motional-narrowing parameter $\alpha_d$. The width of the spectral peak (\ref{eq:slow_noise_simplify}) is given by the bandwidth of the frequency noise $2\Gamma_d$. The peak is non-Lorentzian. However, the sum over $n$ in Eq.~(\ref{eq:slow_noise_simplify}) is dominated by the term $n=1$, and therefore the peak is close to Lorentzian.

\subsection{Evolution of $\Phi_F(\omega)$ with the varying bandwidth and strength of the frequency noise}
\label{subsec:varying_Gamma_d}

\begin{figure}[h]
\includegraphics[width = 6.5truecm]
{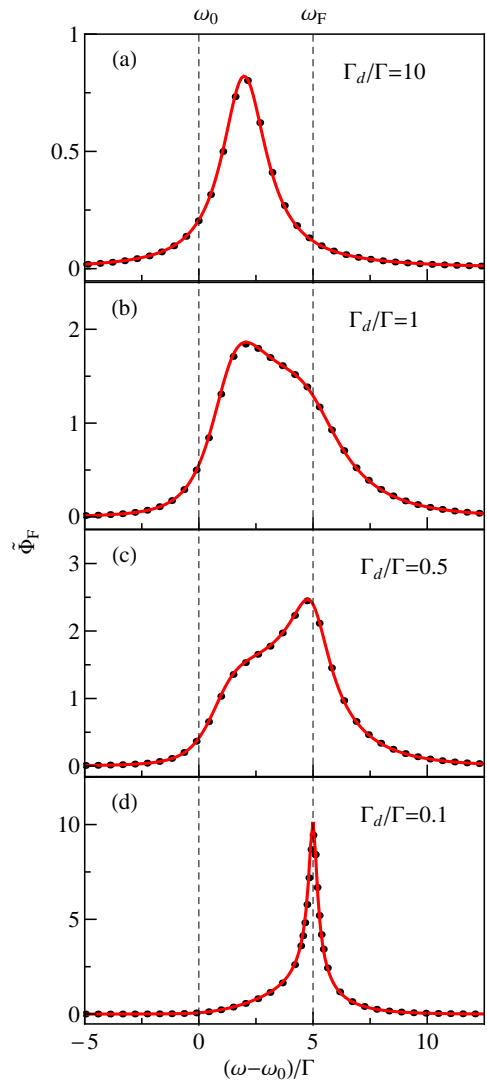}
\caption{The scaled driving-induced part of the power spectrum of the driven mode dispersively coupled to another mode, which we call the $d$-mode. Thermal fluctuations of the $d$-mode lead to frequency fluctuations of the driven mode. Panels (a) to (d) show the change of the spectrum with the varying ratio $\Gamma_d/\Gamma$ of the decay rates of the $d$-mode and the driven mode. The scaled strength (standard deviation) of the frequency noise is $\alpha_d\Gamma_d / \Gamma=1$. The spectrum $\Phi_F(\omega)$ is scaled using the noise-free susceptibility $\chi_0(\omega_F)$, Eq.~(\ref{eq:chi_0}), $\tilde\Phi_F = 4\Gamma\Phi_F/|\chi_0(\omega_F)|^2$. The solid lines and the dots show the analytical theory and the numerical simulations, respectively.}
\label{fig:change_Gamma_d}
\end{figure}

Fig.~\ref{fig:change_Gamma_d} shows the evolution of the driving-induced power spectrum $\Phi_F(\omega)$ with the varying ratio of the decay rates $\Gamma_d/\Gamma$, i.e., the varying ratio  of the bandwidth of the frequency noise  and the decay rate of the driven mode. We use as a scaling factor the susceptibility $\chi_0$ of the driven mode  in the absence of dispersive coupling,
\begin{equation}
\label{eq:chi_0}
\chi_0(\omega_F)=i[2\omega_0(\Gamma-i\delta\omega_F)]^{-1}.
\end{equation}

In Fig.~\ref{fig:change_Gamma_d}~(a), the frequency noise bandwidth is much larger than the width of the spectrum $\Phi_0(\omega)$ in the absence of driving. The spectrum $\Phi_F(\omega)$ is close to a Lorentzian centered near the shifted eigenfrequency of the driven mode, see Eq.~(\ref{eq:broad_band}); for small $\alpha_d$ the shift should be $2\alpha_d\Gamma_d$, whereas the halfwidth should be close to $\Gamma + 2\alpha_d^2\Gamma_d$,\cite{Dykman1971} which agrees with the numerics. In Fig.~\ref{fig:change_Gamma_d}~(d), on the other hand, the noise bandwidth is small. The spectrum $\Phi_F(\omega)$ is a narrow peak near the driving frequency $\omega_F$, see Eq.~(\ref{eq:slow_noise_simplify}), with halfwidth $\approx \Gamma_d$. In Figs.~\ref{fig:change_Gamma_d}~(b)and (c) the frequency noise bandwidth is comparable to the width of the spectrum $\Phi_0(\omega)$. In this case the spectrum $\Phi_F(\omega)$ displays two partly overlapping peaks. The overlapping can be reduced by tuning the driving frequency $\omega_F$ further away from the resonance, see below.

Fig.~\ref{fig:change_amplitude} shows the evolution of the spectrum $\Phi_F(\omega)$ with the varying strength (standard deviation)  $2\alpha_d\Gamma_d$ of the frequency noise. The frequency noise bandwidth $2\Gamma_d$ is chosen to be close to the decay rate of the driven mode $\Gamma$. The driving frequency $\omega_F$ is tuned away from resonance so that the two peaks of $\Phi_F(\omega)$ are well separated. An insight into the shape of the peaks can be gained from the aforementioned similarity of the spectrum $\Phi_F(\omega)$ with the spectrum of fluorescence and quasi-elastic light scattering by a periodically driven oscillating charge. 

For weak frequency noise, curve 1 in Fig.~\ref{fig:change_amplitude} , the peaks are located near $\omega_F$ (quasi-elastic scattering) and $\omega_0$ (fluorescence), cf. Eq.~(\ref{eq:weak_noise}).  As the noise strength increases, the peak near $\omega_0$ becomes broader and the position of its maximum shifts to higher frequency (if $\gamma_d>0$, as assumed in the figure). This resembles the evolution of the spectrum $\Phi_0(\omega)$ in the absence of driving with increasing $\alpha_d$; this evolution is shown in the inset of Fig.~\ref{fig:change_amplitude}. For $\alpha_d > 1$ the peak becomes non-Lorentzian and asymmetric. 

In contrast, the shape of the peak located near $\omega_F$ stays almost  the same with varying noise strength. This is consistent with the picture of quasi-elastic scattering, where the width of the peak is determined by the frequency noise bandwidth. To illustrate how persistent this behavior is, we scaled the spectra in Fig.~\ref{fig:change_amplitude} so that at their maxima at $\omega_F$ the spectra have the same height for different $\alpha_d$.

\begin{figure}[h]
\includegraphics[width = 8truecm]
{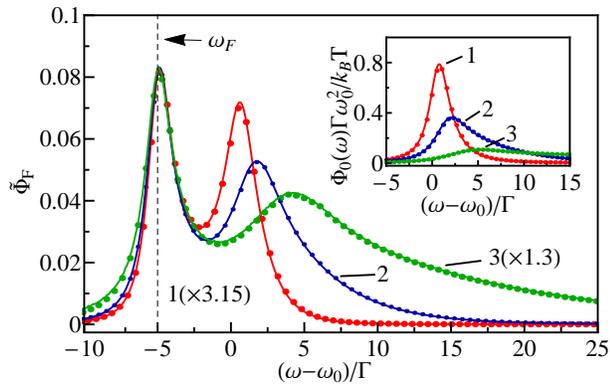}
\caption{The evolution of the driving-induced part of the power spectrum with the varying strength of the frequency noise due to dispersive coupling. Curve 1 to 3 refer to the scaled standard deviation of the noise $\alpha_d\Gamma_d/\Gamma = 0.5$, 2.5, and 12.5, respectively. The ratio of the noise bandwidth to the decay rate of the driven mode is $2\Gamma_d/\Gamma = 1$. The scaled detuning of the driving frequency from the eigenfrequency of the driven mode is $\delta\omega_F/\Gamma = -5$. The spectrum is scaled using the noise-free susceptibility $\chi_0(\omega_F)$, Eq.~(\ref{eq:chi_0}), $\tilde\Phi_F = 4\Gamma\Phi_F/|\chi_0(\omega_F)|^2$. The curves 1 and 3 are additionally scaled by factors 3.15 and 1.3, respectively, so that the peaks near $\omega_F$ have the same height.  The inset shows the spectrum $\Phi_0(\omega)$  in the absence of driving for the same values of the frequency noise strength $\alpha_d\Gamma_d/\Gamma$ as in the main panel. The solid lines and the dots show the analytical theory and the simulations, respectively.}
\label{fig:change_amplitude}
\end{figure} 

\subsection{Effect on $\Phi_F(\omega)$ of the detuning of the driving frequency}
\label{sec:change_detuning}

\begin{figure}[h]
\includegraphics[width = 7truecm]
{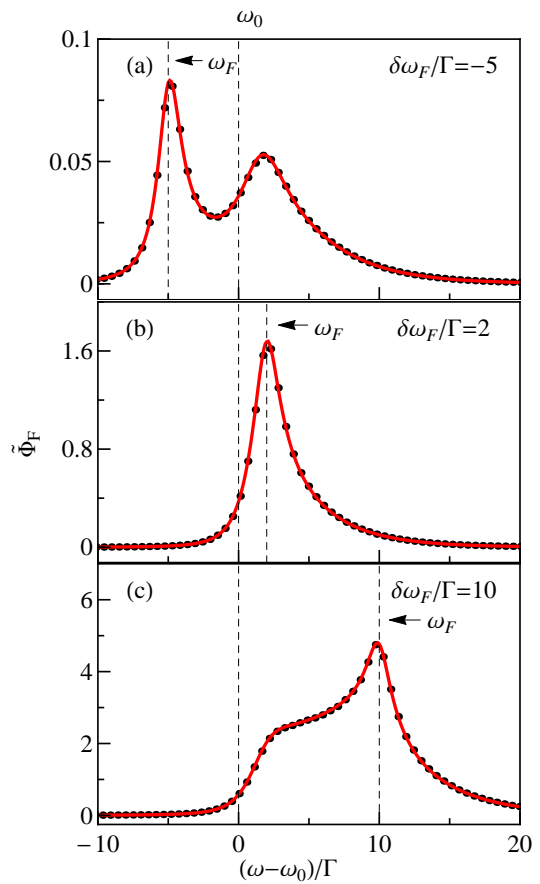}
\caption{The evolution of the driving-induced part of the power spectrum with the varying detuning of the driving frequency $\omega_F$.  The scaled strength of the frequency noise induced by the dispersive coupling is $\alpha_d\Gamma_d/\Gamma =2.5$. The ratio of the noise bandwidth to the decay rate of the driven mode is $2\Gamma_d/\Gamma = 1$.  The spectrum is scaled using the noise-free susceptibility $\chi_0(\omega_F)$, Eq.~(\ref{eq:chi_0}),  $\tilde\Phi_F = 4\Gamma\Phi_F/|\chi_0(\omega_F)|^2$. The solid lines and the dots show the analytical theory and the simulations, respectively.}
\label{fig:change_detuning}
\end{figure}

To provide more insight into the nature of the double-peak structure of the spectrum $\Phi_F(\omega)$ for $\Gamma\sim \Gamma_d$, we show in Fig.~\ref{fig:change_detuning} the effect of detuning of the driving frequency $\omega_F$ from resonance. Panels (a), (b), and (c) refer to the driving frequency being red detuned, equal to, and blue detuned from the the maximum of the spectrum $\Phi_0(\omega)$ in the absence of driving, respectively. The results we show refer to the dispersive coupling constant $\gamma_d>0$. For $\gamma_d<0$, the plots should be mirror-reflected with respect to $\omega-\omega_0$, and  $\omega_F-\omega_0$ should be replaced with $\omega_0-\omega_F$.   

The peak located near the frequency $\omega_F$ is well resolved  in Fig.~\ref{fig:change_detuning}~(a). It moves along with $\omega_F$ as the latter varies. In Fig.~\ref{fig:change_detuning}~(a) one can also see a broader peak, which is located close to $\omega_0$ and essentially does not change its position as $\omega_F$ changes. For small frequency-noise bandwidth, the peak at $\omega_F$ becomes narrow and is described by Eq.~(\ref{eq:slow_noise_simplify}). However, it is well-resolved for large frequency detuning even where the noise bandwidth and the width of the spectrum $\Phi_0(\omega)$ are of the same order of magnitude. If the widths are close and $\omega_F$ is close to resonance, the peaks overlap and cannot be identified, as seen in panel (b). The areas of the peaks are dramatically different for red and blue detuning. This is due to the asymmetry of the spectrum $\Phi_0(\omega)$ in the presence of the frequency noise induced by dispersive coupling, see  the inset of Fig.~(\ref{fig:change_amplitude}). As seen from Fig.~\ref{fig:change_Gamma_d}, for very small $\Gamma_d/\Gamma$ the peak near the oscillator eigenfrequency disappears; this was discussed earlier in the case of weak noise, but is also true in a general case.

\subsection{The area of the driving induced power spectrum}
\label{sec:area}
The area $S_F$ of the driving induced power spectrum $\Phi_F(\omega)$ is defined as $S_F = \int_{0}^{\infty}d\omega \Phi_F(\omega)$. The major contribution to the integral comes from the frequency range where $|\omega - \omega_F|, |\omega-\omega_0|\ll \omega_F$. Then integration over $\omega$ in Eq.~(\ref{eq:Phi_F}) gives a factor $2\pi\delta(t)$.
Further simplification comes from changing from integrating over $t'$ and $t_1'$ to integrating over $t'$ and $t'-t_1'$ and using Eq.~(\ref{eq:susceptibility}) for the susceptibility of the mode. The result reads
\begin{align}
\label{eq:area_simplify}
S_F= \frac{\pi}{4\omega_0\Gamma} {\rm Im} \chi(\omega_F)
-\frac{\pi}{2}|\chi(\omega_F)|^2.
\end{align}
This reduces the calculation of the area $S_F$ just to finding the susceptibility $\chi(\omega_F)$ of the mode. This susceptibility with account taken of the dispersive coupling is given by Eqs.~(\ref{eq:susceptibility}) and (\ref{eq:chi_explicit}). 

The behavior of the area $S_F$ can be found explicitly for small and large $\alpha_d$. In the limit of small $\alpha_d$, where the frequency noise is weak,  from Eq.~(\ref{eq:weak_noise}) 
$S_F \propto \alpha_d^2$. For large $\alpha_d$, it is convenient to write $\tilde\chi(t)$ in Eq.~(\ref{eq:chi_explicit}) as $\tilde\chi(t) \approx (2/\sqrt{i\alpha_d})\sum_{n=0}^{\infty} \exp[-2n(i\alpha_d)^{-1/2}-(2n+1)a_dt]$, where we assumed $\gamma_d>0$; the ultimate result is independent of the sign of $\gamma_d$. The susceptibility $\chi(\omega_F)$ is given by the integral of $\tilde\chi(t)$ over $t$, Eq.~(\ref{eq:susceptibility}). In the limit $\Gamma_d\alpha_d^{1/2}\gg |\Gamma-i\delta\omega_F|$  from Eq.~(\ref{eq:susceptibility}) $\chi(\omega_F)\approx (2\omega_0\alpha_d\Gamma_d)^{-1} \sum \exp[-2n(i\alpha_d)^{-1/2}]/(2n+1)$. To the leading order in $1/\alpha_d$  this gives
\begin{equation}
\label{eq:chi_large_alpha_d}
\chi(\omega_F) \approx \left[\frac{1}{2}\ln (4\alpha_d)  + i\frac{\pi}{4}\right]/4\omega_0\Gamma_d\alpha_d.
\end{equation}
We see from  Eqs.~(\ref{eq:area_simplify}) and (\ref{eq:chi_large_alpha_d}) that $S_F \propto \alpha_d^{-1}$ falls down with increasing $\alpha_d$ for large $\alpha_d$.

\begin{figure}[h]
\includegraphics[width = 7truecm]
{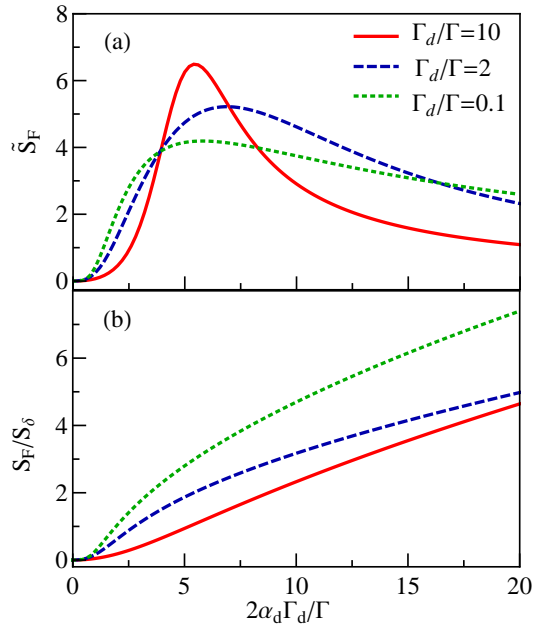}
\caption{The area of the driving-induced part of the power spectrum as a function of $\alpha_d$ for different ratio of the frequency noise bandwidth to the decay rate of the driven mode $2\Gamma_d/\Gamma$. The red (solid), blue (dashed), and green (dotted) lines refer to $\Gamma_d/\Gamma =$ 10, 2, and 0.1, respectively. The relative detuning of the driving frequency is $\delta\omega_F/\Gamma = 5$. In panel (a), the area $S_F$ is scaled using the noise-free susceptibility $\chi_0(\omega_F)$,  Eq.~(\ref{eq:chi_0}),  $\tilde S_F = 2S_F/\pi |\chi_0(\omega_F)|^2$. In panel (b), $S_F$ is scaled by the area of the $\delta$ peak in the power spectrum of the driven mode, $S_\delta = \pi|\chi(\omega_F)|^2/2$. }
\label{fig:area}
\end{figure}

The nonmonotonic dependence of the area $S_F$ on the parameter $\alpha_d$, which is expected from the above asymptotic expressions, is indeed seen in Fig.~\ref{fig:area}(a). This figure shows  the area $S_F$ as a function of the frequency noise strength $\alpha_d\Gamma_d$ for different $\Gamma_d/\Gamma$. The position of the maximum of $S_F$ sensitively depends on $\Gamma_d/\Gamma$.

In terms of a comparison with experiment, it is advantageous to scale the spectrum $\Phi_F$, and in particular the area $S_F$,  by the area of the $\delta$-peak in the power spectrum of the driven mode. This area is given by the expression $S_\delta =  (\pi/2)|\chi(\omega_F)|^2$, cf. Eq.~(\ref{eq:correlator_defined}). The quantities measured in the experiment are $F^2S_F$ and $F^2S_\delta$. The unknown scaled field intensity $F^2$ drops out from their ratio. From Eqs.~(\ref{eq:area_simplify}) and (\ref{eq:chi_large_alpha_d})  $S_F/S_\delta \propto \alpha_d/\ln^2\alpha_d$ increases with $\alpha_d$ for large $\alpha_d$. For small $\alpha_d$,  $S_F/S_\delta \propto \alpha_d^2$ also increases with $\alpha_d$. On the whole, we found that $S_F/S_\delta $ monotonically increases with $\alpha_d$. This increase is seen in Fig.~\ref{fig:area}~(b).

\section{Power spectrum of a driven nonlinear oscillator}
\label{sec:internal_nonlinearity}

An important contribution to the broadening of the spectra of mesoscopic oscillators can come from their internal nonlinearity.\cite{Dykman2012b} The vibration frequency of a nonlinear oscillator depends on the vibration amplitude. Therefore thermal fluctuations of the amplitude lead to frequency fluctuations. The analysis of the spectra is complicated by the interplay of the frequency fluctuations that come from the amplitude fluctuations and the frequency uncertainty that comes from the oscillator decay. Nevertheless the linear susceptibility could be found for an arbitrary relation between the standard deviation of the frequency $\Delta\omega$ and the decay rate $\Gamma$.\cite{Dykman1971}. The power spectrum of a nonlinear oscillator in the absence of driving is generally asymmetric and non-Lorentzian.

Finding the driving-induced terms in the power spectrum is  still more complicated. The oscillator displacement is nonlinear in the driving field amplitude $F$, and the driving-induced part of the power spectrum $\Phi(\omega)$ is not quadratic in $F$. However, if the field is weak,  Eq.~(\ref{eq:correlator_defined}) for $\Phi(\omega)$ applies. In the calculation of $\Phi_F(\omega)$  one should take into account terms in the oscillator displacement that are quadratic in $F$, which is generic for nonlinear systems.\cite{Zhang2014a} 

We assume that the nonlinear part of the oscillator energy is small compared to the linear part. Then the nonlinear term in the oscillator energy can be taken in the form of $\gamma q^4/4$.\cite{LL_Mechanics2004} The oscillator equation of motion in the rotating wave approximation is similar to Eq.~(\ref{eq:eom}), except that it now contains the term due to the internal nonlinearity,
\begin{equation}
\label{eq:dot_u_nonlinear}
\dot u =  -(\Gamma + i\delta\omega_F)u +\frac{3i\gamma}{2\omega_0}|u|^2u-i \frac{F}{4\omega_0} + f(t).
\end{equation}
In this section we do not discuss the effect of dispersive coupling, and the frequency noise that comes from this coupling is not included into Eq.~(\ref{eq:dot_u_nonlinear}).

To find $\Phi_F(\omega)$, we first consider the dynamics of a driven nonlinear oscillator without fluctuations and then take fluctuations into account. The stationary solution $u_{\rm st}$ of Eq.~(\ref{eq:dot_u_nonlinear}) in the absence of the noise $f(t)$ can be found by setting $\dot u = 0$. For weak driving, $u_{\rm st}$ is a series in $F$, which contains only odd powers of $F$. Since we are interested in the terms which are linear or quadratic in $F$, it is sufficient to keep only the leading term, $u_{\rm st} = F/4i\omega_0(\Gamma+i\delta\omega_F)$. One then substitutes into Eq.~(\ref{eq:dot_u_nonlinear}) $u(t)=u_{\rm st} + \delta u(t)$. The deviation $\delta u(t)$ is due only to the noise,
\begin{align}
\label{eq:dot_delta_u}
\delta \dot u = &- (i\delta\omega_F  + \Gamma) \delta u \nonumber + \frac{3i\gamma}{2\omega_0} \left(|\delta u|^2 \delta u + 2u_{\rm st}|\delta u|^2 + u_{\rm st}^* \delta u^2\right. \nonumber  \\
&\left.+ 2|u_{\rm st}|^2 \delta u + u_{\rm st}^2 \delta u^*\right) + f(t).
\end{align}
Time evolution of $\delta u(t)$ depends on the driving field in terms of $u_{\rm st}$. We find this time evolution in the two limiting cases.

\subsection{Weak nonlinearity}

The analysis of the dynamics simplifies in the case of small nonlinearity-induced spread of the oscillator frequency $\Delta\omega$ compared to the decay rate $\Gamma$. As seen from Eq.~(\ref{eq:dot_u_nonlinear}), in the absence of driving the frequency shift is quadratic in the vibration amplitude $\propto |u|^2$,\cite{LL_Mechanics2004}, and therefore the frequency spread is determine by the standard deviation of $|u|^2$ due to the thermal noise. This gives $\Delta\omega =  3|\gamma|k_BT/8\omega_0^3$.  

For $\Delta \omega \ll \Gamma$, it is sufficient to keep only the linear in $\delta u$ terms in Eq.~(\ref{eq:dot_delta_u}).\cite{Dykman1979a,Drummond1980c} A straightforward calculation then gives a simple expression for the the driving-induced power spectrum,
\begin{align}
\label{eq:weak_nonlinearity}
\Phi_F(\omega)\approx \frac{3\gamma k_BT}{8\omega_0^5}\frac{(\omega - \omega_0)\Gamma}{(\Gamma^2+\delta\omega_F^2) [\Gamma^2+(\omega-\omega_0)^2]^2}.
\end{align}

The spectrum (\ref{eq:weak_nonlinearity}) is proportional to the derivative of the Lorentzian spectrum of the harmonic oscillator $\Phi_0(\omega)\propto 1/[\Gamma^2 + (\omega-\omega_0)^2]$ over $\omega$. It has a characteristic dispersive shape, being of the opposite signs on the other sides of $\omega _0$. This is the result of the shift of the oscillator vibration frequency $\propto \gamma F^2$ due to the driving. Such shift is the main effect of the driving for small $\Delta\omega/\Gamma$.

\subsection{Large detuning of the driving field frequency}

For arbitrary $\Delta\omega/\Gamma$, the analysis is simplified if the detuning of the driving field frequency from the small-amplitude oscillator frequency $|\delta\omega_F| \gg \Gamma, \Delta\omega$. In this case, one can change variables in Eq.~(\ref{eq:dot_delta_u}) to $\delta\tilde u(t) = \delta u(t) e^{i\delta\omega_F t}$. The right-hand side of the resulting equation for $\delta\tilde u$, besides the noise term, has terms that smoothly depend on time on the scale $|\delta\omega_F|^{-1}$ and terms that oscillate as $\exp(\pm i\delta\omega_Ft), \exp(2i\delta\omega_Ft)$. These oscillating terms can be considered a perturbation. To the first order of the perturbation theory, the equation for the smooth terms takes the form
\begin{align}
\label{eq:dot_delta_u_bar}
&\delta\dot{\tilde u} = - \Gamma \delta\tilde u   + \frac{3i\gamma}{\omega_0}|u_{\rm st}|^2 \delta\tilde u \nonumber\\
& + \left(1+ \frac{9\gamma |u_{\rm st}|^2}{\omega_0\delta\omega_F}\right)  \frac{3i\gamma}{2\omega_0} |\delta\tilde u|^2\delta\tilde u + \tilde f(t),
\end{align}
where $\tilde f(t) = f(t)e^{i\delta\omega_F t}$.  We keep in this equation the terms $\propto |u_{\rm st}|^2\propto F^2$. These terms contribute to the spectrum $\Phi_F(\omega)$. The terms of higher oder in $|u_{\rm st}|^2$ have been discarded.

Equation~(\ref{eq:dot_delta_u_bar}) has the same form as the equation of motion for the complex amplitude $u(t)$ in the absence of driving, i.e., Eq.~(\ref{eq:dot_u_nonlinear}) with $F = 0$. The noise $\tilde f(t)$ has the same correlation function as $f(t)$. Therefore the power spectrum of $\delta\tilde u(t)$ is the same as the power spectrum of a nonlinear oscillator found earlier,\cite{Dykman1971} with the renormalized parameters: the eigenfrequency is shifted by $3\gamma |u_{\rm st}|^2/\omega_0$ and the nonlinearity parameter is multiplied by the factor $1+9\gamma|u_{\rm st}|^2/\omega_0\delta\omega_F$. We note that the correction $\propto |u_{\rm st}|^2$ in this factor, which comes from the perturbation theory in $1/\delta\omega_F$, is small. 

To find $\Phi_F(\omega)$ we have to expand the result\cite{Dykman1971} with the appropriately renormalized parameters  to the first order in $|u_{\rm st}|^2$. This gives 
\begin{align}
\label{eq:nonlinear_large_detuning}
&F^2\Phi_F(\omega) = \beta\{\partial_\beta[\Phi_0(\omega - 2\beta\delta\omega_F;\Delta\omega(1+6\beta)) ] \}_{\beta = 0}, \nonumber\\
&\Phi_0(\omega;\Delta\omega)=\frac{k_BT}{\omega_0^2}\Re \int_0^{\infty}dt \exp\{[i(\omega-\omega_0)+\Gamma] t\}\nonumber\\
&\times [\cosh(at) + (\Gamma/a)(1+2i\alpha\sgn\gamma)\sinh(at)]^{-2}.
\end{align}
The parameters $\alpha$ and $a$ have the same structure and the same physical meaning as the parameters $\alpha_d$ and $a_d$ used before, $\alpha = \Delta\omega/\Gamma$ and $a = \Gamma(1+4i\alpha\sgn\gamma)^{1/2}$, whereas $\beta = 3\gamma F^2/32\omega_0^3(\delta\omega_F)^3$ is the scaled intensity of the driving field.

The major contribution to $\Phi_F(\omega)$ as given by Eq.~(\ref{eq:nonlinear_large_detuning}) for large $|\delta\omega_F|/\Delta\omega$ comes from the frequency shift of the spectrum without driving $\Phi_0(\omega)$ and is determined by $-2\delta\omega_F \partial_\omega\Phi_0(\omega; \Delta\omega)$. 
Physically, this results again corresponds to the shift of the oscillator eigenfrequency associated with the forced vibrations, and the spectrum $\Phi_F$ again has the characteristic shape of a dispersive curve. 
To the next order in $1/\delta\omega_F$, the driving broadens or narrows the spectrum depending on the sign of $\gamma/\delta\omega_F$ by renormalizing the nonlinearity-induced standard deviation of the oscillator frequency $\Delta\omega$.

\begin{figure}[h]
\includegraphics[width = 8truecm]
{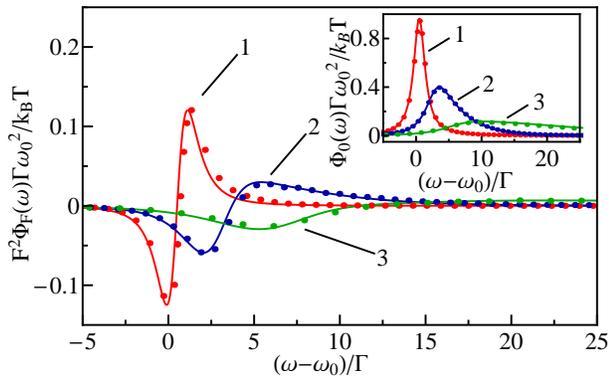}
\caption{The driving-induced part of the power spectrum of a nonlinear oscillator for large detuning of the driving frequency, $\delta\omega_F/\Delta\omega = 40$. The solid curves and the dots show the analytical expressions and the results of simulations, respectively. The values of the nonlinearity parameter and the scaled driving strength for the curves 1 to 3 are, respectively, $\alpha\equiv \Delta\omega/\Gamma =$~0.125, 1.25, and 5, and $\beta\equiv 3\gamma F^2/32\omega_0^3\delta\omega_F^3 =$~ 0.016, 0.004, and 0.004. The inset shows the change of the power spectrum in the absence of driving with varying $\Delta\omega/\Gamma$. }
\label{fig:nonlinear_large_detuning}
\end{figure}

\begin{figure}[]
\includegraphics[width = 8truecm]
{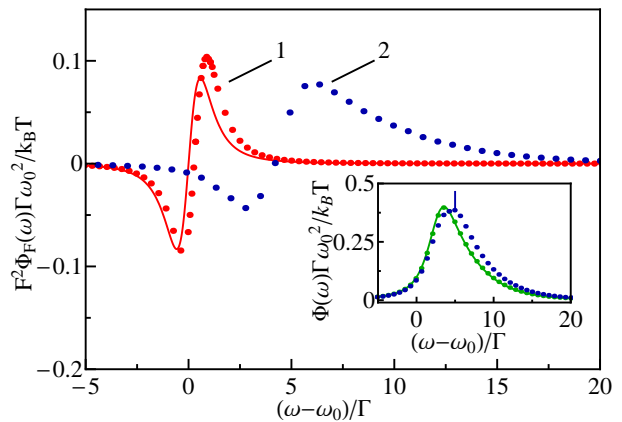}
\caption{The driving-induced part of the spectrum of a nonlinear oscillator for small detuning of the driving frequency. The solid curve (red) shows the analytical results for $\Phi_F(\omega)$ for small $\Delta\omega/\Gamma$ for the same parameters as the dotted curve 1. The dots show the results of simulations.  The scaled values of the nonlinearity parameter, the detuning, and the driving strength on the curves  1 and 2 are, respectively, $\alpha\equiv \Delta\omega/\Gamma =$~0.05, and 1.25, $\delta\omega_F/\Gamma =$~0.5 and 5, and $\beta\equiv 3\gamma F^2/32\omega_0^3(\delta\omega_F)^3 =$~0.64 and 0.01. The inset shows the full spectrum for the parameters of curve 2 (blue dots, simulations); the spectrum without driving for the same $\Delta\omega/\Gamma$ is shown by the solid line (analytical) and (green) dots on top of this line, which are obtained by simulations.}
\label{fig:nonlinear_small_detuning}
\end{figure}

\subsection{Numerical simulations}

The analytical results on the spectra of the modulated nonlinear oscillator, Eq.~(\ref{eq:nonlinear_large_detuning}), are compared with the results of numerical simulations in Fig.~\ref{fig:nonlinear_large_detuning}. The spectrum $\Phi_F(\omega)$ generally has a positive and negative parts, in a dramatic distinction from the case of a linear oscillator dispersively coupled to another oscillator. As $\Delta\omega/\Gamma$ increases, the shape of $\Phi_F(\omega)$ becomes more complicated, in particular, the positive and negative parts become asymmetric. 

The simulations were performed in the same way as for the dispersively coupled modes by integrating the stochastic differential equations (\ref{eq:dot_u_nonlinear}). We verified that the values of the modulating field amplitude $F$ were in the range where the driving-induced term in the power spectrum was quadratic in $F$. As seen from this figure, the simulations are in excellent agreement with the analytical results. 

In the intermediate range, where the nonlinearity is not weak and the driving is not too far detuned, i.e., $|\delta\omega_F|\sim \max(\Gamma, \Delta\omega)$, we obtained the spectrum $\Phi_F(\omega)$ by running numerical simulations. These results are presented in Fig.~\ref{fig:nonlinear_small_detuning}.  They show that the general trend seen in Fig.~\ref{fig:nonlinear_large_detuning} that $\Phi_F(\omega)$ changes signs and is asymmetric for a nonlinear oscillator persists in this case as well.

\section{Conclusions}

In terms of experimental studies of mesoscopic vibrational systems, the major result of this paper is the suggestion of a way to single out and characterize the dispersive (nonresonant) coupling between vibrational modes. We have shown that this coupling leads to a specific, generally double-peak extra structure in the power spectrum of a mode when this mode is driven close to resonance. The dispersive-coupling induced part of the power spectrum is quadratic in the driving field amplitude. It varies significantly with the detuning of the driving frequency from the mode eigenfrequency. 

The "tune off to read off" approach allows one to study separately two effects by changing the driving frequency. One is the dispersive-coupling induced broadening of the spectral peak of the linear response, which is of significant interest for mesoscopic modes.\cite{Barnard2012,Venstra2012,Matheny2013,Miao2014,Vinante2014} The other is the decay of the ``invisible" mode that is dispersively coupled to the studied mode but may not be necessarily directly accessible. The double-peak structure of the driving-induced power spectrum sensitively depends both on the strength of the dispersive coupling and the mode parameters.

Another important feature of the driving-induced spectrum is the qualitative difference between the effects of nonlinear dispersive coupling to other modes and the internal nonlinearity of the studied mode. Both nonlinearities are known to broaden, in a somewhat similar way,\cite{Dykman1971} the linear response spectrum in the presence of thermal fluctuations. However, in the case of internal nonlinearity, the driving-induced part of the power spectrum changes sign as a function of frequency,  i.e., it has peaks of the opposite signs and is similar (and is close, in a certain parameter range) to the derivative of the power spectrum without driving.

In terms of the theory, the paper describes a path-integral method that enables finding in an explicit form the spectrum of a driven oscillator in the presence of non-Gaussian fluctuations of its frequency, which result from dispersive coupling to other modes. The results apply for an arbitrary ratio between the relevant parameters of the system. These parameters are  the magnitude (standard deviation) of the frequency fluctuations $\Delta\omega$, their reciprocal correlation time, which is given by the decay rate of the dispersively-coupled mode that causes the fluctuations, the decay rate of the driven mode itself, and the detuning of the driving frequency. It is the presence of several parameters that makes it complicated to identify the broadening mechanisms from the linear response spectra. The results of the paper show the qualitative difference between the effects of these parameters on the power spectrum when the oscillator is driven. This enables their identification.

The results are easy to extend to the case of dispersive coupling to several modes. The contributions of different modes to the frequency fluctuations of the studied mode, and therefore to the random accumulation of its phase, are additive and mutually independent. Then the averaging over the phase accumulation in Eq.~(\ref{eq:averaging_object}) can be done independently for each of them. The result is the product of the averages [functions $G(t,t',t_1')$] calculated for each mode taken separately, with the appropriate coupling parameters and the decay rates of the modes. 

Generally, in nanomechanical systems the internal (Duffing) and dispersive nonlinearities can be of the same order of magnitude. If the studied mode has a much higher frequency than the mode to which it is dispersively coupled, its fluctuations can be comparatively weaker making the effect of the dispersive coupling stronger. Also if there are several modes dispersively coupled to the mode of interest, their cumulative effect can be stronger than the effect of the internal nonlinearity. This makes it even more important to be able to distinguish the effects, which the proposed approach suggests.

The results  immediately extend to the parameter range where the driven mode has high frequency and is in the quantum regime, $\hbar\omega_0 > k_BT$. For dispersive coupling to a classical mode, the driving-induced part of the power spectrum is described by the same expression as where the driven mode is also classical. This case is of particular interest for optomechanics, where the high-frequency optical cavity mode can be dispersively coupled to a low-frequency mechanical mode.\cite{Sankey2010,Aspelmeyer2014,Ludwig2012} Driving the cavity mode leads in this case to a characteristic radiation described by this paper.

\acknowledgements
This research was supported in part by the U.S. Army Research Office (W911NF-12-1-0235) and the National Science Foundation (DMR-1514591).


\appendix

\section{The transfer-matrix type construction}
\label{sec:transfer_matrix}

The central part of the calculation of the driving-induced power spectrum is the averaging over the frequency noise due to dispersive coupling. Equations (\ref{eq:averaging_object}) and (\ref{eq:G_final}) reduce this averaging to solving an ordinary differential equation (\ref{eq:ddot_determinant}) with the coefficient that varies with time stepwise. The solution can be simplified by taking advantage of this specific time dependence. 

From Eq.~(\ref{eq:k_limits}), the interval $(t_0,t)$ in Eq.~(\ref{eq:ddot_determinant}) is separated into three regions $m=1,2,3$ within which the time-dependent coefficient $k(t'')= \bar k_m$ is constant. The boundaries between the regions $\tau$ and $\tau'$ and the values of $\bar k_m$ are specified in  Eq.~(\ref{eq:k_limits}). We enumerate the regions in the order of decreasing time, that is, the region $\tau<t''<t$ corresponds to $m=1$, etc. In each region
\begin{align}
\label{eq:solution_D}
&D(t'';k)= M_{11}(t''-t_0;m) A_m + M_{12}(t''-t_0;m) B_m.
\end{align}
Here, $M_{ij}$ are the matrix elements of the matrix
\begin{align}
\label{eq:M_matrix}
&\hat M(t''; m) = 
\begin{pmatrix}
\cosh a_m t'' & \sinh a_m t'' \\
a_m \sinh a_m t'' & a_m \cosh a_m t''  
\end{pmatrix}
, \nonumber\\
&a_m\equiv a(\bar k_m)=\Gamma_d(1-4i\alpha_d\bar k_m)^{1/2}
\end{align}
We note that, from Eq.~(\ref{eq:k_limits}), $a_1 = \Gamma_d(1+ 4i\alpha_d\sgn\gamma_d)^{1/2}$, whereas $a_2 = \Gamma_d$; $a_3$ is equal to either $a_1$ or $a_1^*$ depending on whether $t'<t_1'$ or $t' > t_1'$ in the argument of the $G$-function in (\ref{eq:G_final}).

The values of $A_1,B_1$ in Eq.~(\ref{eq:solution_D}) are determined by the conditions  $D(t;k) =1, \dot D(t;k)=-\Gamma_d$.  The values of $A_m,B_m$ for $m=2,3$ are found from the continuity of $D(t'';k), \dot D(t'';k)$ at the boundaries $t''=\tau, \tau'$. 

Function $G(t,t',t_1')$ in Eq.~(\ref{eq:G_final}) is determined by $D(t_0;k)=A_3$ and $\dot D(t_0;k) = a_3B_3$. From 
Eqs.~(\ref{eq:solution_D}) and (\ref{eq:M_matrix}) we have
\begin{align}
\label{eq:coefficient}
&\begin{pmatrix}
A_3 \\ B_3 
\end{pmatrix}
 =\hat M^{-1}(\tau'-t_0; 3)\hat M(\tau'-t_0; 2)\hat M^{-1}(\tau -t_0; 2) \nonumber \\
&\times\hat M(\tau -t_0; 1)\hat M^{-1}(t-t_0; 1) \begin{pmatrix} 1 \\ -\Gamma_d \end{pmatrix} ,
\end{align}
This simple relation combined with Eq.~(\ref{eq:G_final}) give the integrand in the expression for the power spectrum $\Phi_F(\omega)$ in a simple form, which is convenient for numerical integration. The expression (\ref{eq:coefficient}) can be evaluated in the explicit form. The result is given in Sec.~\ref{sec:app_G_explicit}. It is advantageous when one looks for the asymptotic expressions for the spectrum $\Phi_F(\omega)$.

\section{Alternative path-integral approach to averaging over frequency noise}
\label{sec:app_DK_method}

Here we provide an alternative approach to evaluating function $G(t,t',t_1')$, which is defined by Eq.~(\ref{eq:G_function}) and describes the outcome of averaging over the frequency noise. The method is related, albeit fairly remotely, to the method developed for calculating the power spectrum of a nonlinear oscillator in the absence of driving.\cite{Dykman1971,DK_review84} We start with writing  the probability density functional of the Gaussian process $Q_d(t)$ on the whole time axes, $-\infty < t <\infty$, in terms of the correlation function $A(t_2,t_3) = \langle Q_d(t_2)Q_d(t_3)\rangle$ and its inverse $A^{-1}(t_2,t_3)$,
\begin{align}
\label{eq:prob_dist_functional}
&{\cal P}[Q_d(t)] = \exp\left[-\frac{1}{2}\int\int dt_1dt_2 Q_d(t_1)A^{-1}(t_1,t_2)Q_d(t_2)\right], \nonumber\\
&\int dt_2 A^{-1}(t_1,t_2) A(t_2,t_3) = \delta(t_1-t_3), 
\end{align}
 cf. Ref.~\onlinecite{FeynmanQM}). 

From  Eq.~(\ref{eq:G_function}), function $G$ and its derivative $\partial G/ \partial t$ can be written as
\begin{align}
\label{eq:G_via_tilde_P}
&G(t,t',t_1') = \frac{\int {\cal D}[Q_d] \tilde {\cal P}[Q_d]}{ \int {\cal D}[Q_d] {\cal P}[Q_d]}, \nonumber\\  
&\frac{\partial G}{\partial t} = -i\sgn(\gamma_d)\frac{\int {\cal D}[Q_d]Q_d^2(t) \tilde {\cal P}[Q_d]}{ \int {\cal D}[Q_d] {\cal P}[Q_d]}, 
\end{align}
where functional $\tilde {\cal P}$ has the form
\begin{align}
\label{eq:tilde_P_defined}
&\tilde{\cal P}[Q_d] = \exp\left[-\frac{1}{2}\int\int dt_1dt_2 Q_d(t_1)\tilde A^{-1}(t_1,t_2)Q_d(t_2)\right], \nonumber \\
&\tilde A^{-1}(t_1,t_2) = A^{-1}(t_1,t_2)-2ik(t_2)\delta(t_1-t_2) .
\end{align}
Here $k(t_2)$ is a stepwise function, which is equal to 0 or $\pm 1$ in the time interval $(t_0,t)$, where $t_0\equiv \min(t',t_1')$ is defined in Eq.~(\ref{eq:k_limits}). This definition has to be extended in the present formulation, $k(t_2)=0$ for $t_2>t$ and $t_2<t_0$.

A key observation is that functional $\tilde{\cal P}[Q_d]$ is also Gaussian. One can introduce an operator $\tilde A(t,t_1)$ reciprocal to $\tilde A^{-1}$,
\begin{equation}
\label{eq:tilde_A_define}
\int dt_1 \tilde A(t,t_1) \tilde A^{-1}(t_1,t_2) = \delta(t-t_2).
\end{equation}
In terms of this operator, 
\begin{align}
\label{eq:tilde_A_t_t}
 \tilde A(t,t)  = \frac{\int {\cal D}[Q_d] Q_d^2(t) \tilde {\cal P}[Q_d]}{ \int {\cal D}[Q_d] \tilde {\cal P}[Q_d]}
\end{align}
where $\tilde A$ is related to $\tilde A^{-1}$ through
\begin{equation}
\label{eq:tilde_A_define}
\int dt_1 \tilde A(t,t_1) \tilde A^{-1}(t_1,t_2) = \delta(t-t_2).
\end{equation}

Multiplying equation~(\ref{eq:tilde_P_defined}) for $\tilde A^{-1}$ by $A(t_2,t_3)\tilde A(t_1,t)$ and integrating with respect to $t_1, t_2$, we obtain an integral equation for $\tilde A(t,t_3)$,
\begin{align}
\label{eq:integral_tilde_A}
\tilde A(t,t_3) - 2i\int k(t_1) \tilde A(t,t_1) A(t_1,t_3)dt_1 = A(t,t_3).
\end{align}
This equation can be reduced to a differential equation by differentiating twice with respect to $t_3$,
\begin{align}
\label{eq:ddot_tilde_A}
&\frac{\partial^2 \tilde A(t,t_3)}{\partial t_3^2} - \Gamma_d^2[1-4i\alpha_d k(t_3)]\tilde A(t,t_3) = -2\alpha_d\Gamma_d^2\delta(t_3-t).
\end{align}
Interestingly, Eq.~(\ref{eq:ddot_tilde_A}) has the same structure as the differential equation for the "time-dependent" determinant found in the other method, see Eq.~(\ref{eq:ddot_determinant}). Thus it can be solved in a similar fashion as in Appendix~(\ref{sec:transfer_matrix}). The boundary conditions are $\tilde A(t,\pm \infty) = 0$. It follows from the decay of correlations of $Q_d(t)$ . At the values of $t_3$ where $k(t_3)$ changes stepwise, see Eq.~(\ref{eq:k_limits}), $\tilde A$ and $\partial \tilde A /\partial t_3$ remain continuous, except $t_3 = t$, where $\partial \tilde A/\partial t_3$ changes by $2\alpha_d\Gamma_d^2$,as seen from Eq.~(\ref{eq:ddot_tilde_A}).

We can now write Eq.~(\ref{eq:tilde_P_defined}) for function $G(t,t',t_1')$, in terms of function $\tilde A$,
\begin{equation}
\label{eq:dot_G}
\partial_t G(t,t',t_1') = -i\sgn(\gamma_d)\tilde A(t,t) G.
\end{equation}
The boundary condition for this equation is $G(t,t,0)=1$. From the explicit expression for $G$ we also have 
\begin{align}
\label{eq:dot_G_t'}
&\partial_{t'} G(t,t',t_1') = i\sgn(\gamma_d)\tilde A(t',t') G(t,t',t_1'),\nonumber\\
&\partial_{t_1'} G(t,t',t_1') = -i\sgn(\gamma_d)\tilde A(t_1',t_1') G(t,t',t_1').
\end{align}
The solution of these equations reads
\begin{align}
\label{eq:answer_new}
G(t,t',t_1') =& \exp\left\{-i\sgn(\gamma_d)\left[\int_{t'}^t \tilde A(t'',t'')dt''\right.\right.\nonumber\\
&\left.\left. -\int_{t_1'}^0 \tilde A(t'',t'')dt''\right]\right\}.
\end{align}

We have checked that the expression for function $G$ that follows from Eqs.~(\ref{eq:ddot_tilde_A}) and (\ref{eq:answer_new})  coincides with the result obtained in the main text.

\bibliographystyle{apsrev4-1}

\begin{thebibliography}{39}%
\makeatletter
\providecommand \@ifxundefined [1]{%
 \@ifx{#1\undefined}
}%
\providecommand \@ifnum [1]{%
 \ifnum #1\expandafter \@firstoftwo
 \else \expandafter \@secondoftwo
 \fi
}%
\providecommand \@ifx [1]{%
 \ifx #1\expandafter \@firstoftwo
 \else \expandafter \@secondoftwo
 \fi
}%
\providecommand \natexlab [1]{#1}%
\providecommand \enquote  [1]{``#1''}%
\providecommand \bibnamefont  [1]{#1}%
\providecommand \bibfnamefont [1]{#1}%
\providecommand \citenamefont [1]{#1}%
\providecommand \href@noop [0]{\@secondoftwo}%
\providecommand \href [0]{\begingroup \@sanitize@url \@href}%
\providecommand \@href[1]{\@@startlink{#1}\@@href}%
\providecommand \@@href[1]{\endgroup#1\@@endlink}%
\providecommand \@sanitize@url [0]{\catcode `\\12\catcode `\$12\catcode
  `\&12\catcode `\#12\catcode `\^12\catcode `\_12\catcode `\%12\relax}%
\providecommand \@@startlink[1]{}%
\providecommand \@@endlink[0]{}%
\providecommand \url  [0]{\begingroup\@sanitize@url \@url }%
\providecommand \@url [1]{\endgroup\@href {#1}{\urlprefix }}%
\providecommand \urlprefix  [0]{URL }%
\providecommand \Eprint [0]{\href }%
\providecommand \doibase [0]{http://dx.doi.org/}%
\providecommand \selectlanguage [0]{\@gobble}%
\providecommand \bibinfo  [0]{\@secondoftwo}%
\providecommand \bibfield  [0]{\@secondoftwo}%
\providecommand \translation [1]{[#1]}%
\providecommand \BibitemOpen [0]{}%
\providecommand \bibitemStop [0]{}%
\providecommand \bibitemNoStop [0]{.\EOS\space}%
\providecommand \EOS [0]{\spacefactor3000\relax}%
\providecommand \BibitemShut  [1]{\csname bibitem#1\endcsname}%
\let\auto@bib@innerbib\@empty
\bibitem [{\citenamefont {Barnard}\ \emph {et~al.}(2012)\citenamefont
  {Barnard}, \citenamefont {Sazonova}, \citenamefont {van~der Zande},\ and\
  \citenamefont {McEuen}}]{Barnard2012}%
  \BibitemOpen
  \bibfield  {author} {\bibinfo {author} {\bibfnamefont {A.~W.}\ \bibnamefont
  {Barnard}}, \bibinfo {author} {\bibfnamefont {V.}~\bibnamefont {Sazonova}},
  \bibinfo {author} {\bibfnamefont {A.~M.}\ \bibnamefont {van~der Zande}}, \
  and\ \bibinfo {author} {\bibfnamefont {P.~L.}\ \bibnamefont {McEuen}},\
  }\href@noop {} {\bibfield  {journal} {\bibinfo  {journal} {PNAS}\ }\textbf
  {\bibinfo {volume} {109}},\ \bibinfo {pages} {19093} (\bibinfo {year}
  {2012})}\BibitemShut {NoStop}%
\bibitem [{\citenamefont {Eichler}\ \emph {et~al.}(2012)\citenamefont
  {Eichler}, \citenamefont {del {\AA}lamo~Ruiz}, \citenamefont {Plaza},\ and\
  \citenamefont {Bachtold}}]{Eichler2012}%
  \BibitemOpen
  \bibfield  {author} {\bibinfo {author} {\bibfnamefont {A.}~\bibnamefont
  {Eichler}}, \bibinfo {author} {\bibfnamefont {M.}~\bibnamefont {del
  {\AA}lamo~Ruiz}}, \bibinfo {author} {\bibfnamefont {J.~A.}\ \bibnamefont
  {Plaza}}, \ and\ \bibinfo {author} {\bibfnamefont {A.}~\bibnamefont
  {Bachtold}},\ }\href@noop {} {\bibfield  {journal} {\bibinfo  {journal}
  {Phys. Rev. Lett.}\ }\textbf {\bibinfo {volume} {109}},\ \bibinfo {pages}
  {025503} (\bibinfo {year} {2012})}\BibitemShut {NoStop}%
\bibitem [{\citenamefont {Westra}\ \emph {et~al.}(2010)\citenamefont {Westra},
  \citenamefont {Poot}, \citenamefont {van~der Zant},\ and\ \citenamefont
  {Venstra}}]{Westra2010}%
  \BibitemOpen
  \bibfield  {author} {\bibinfo {author} {\bibfnamefont {H.~J.~R.}\
  \bibnamefont {Westra}}, \bibinfo {author} {\bibfnamefont {M.}~\bibnamefont
  {Poot}}, \bibinfo {author} {\bibfnamefont {H.~S.~J.}\ \bibnamefont {van~der
  Zant}}, \ and\ \bibinfo {author} {\bibfnamefont {W.~J.}\ \bibnamefont
  {Venstra}},\ }\href@noop {} {\bibfield  {journal} {\bibinfo  {journal} {Phys.
  Rev. Lett.}\ }\textbf {\bibinfo {volume} {105}},\ \bibinfo {pages} {117205}
  (\bibinfo {year} {2010})}\BibitemShut {NoStop}%
\bibitem [{\citenamefont {Castellanos-Gomez}\ \emph {et~al.}(2012)\citenamefont
  {Castellanos-Gomez}, \citenamefont {Meerwaldt}, \citenamefont {Venstra},
  \citenamefont {van~der Zant},\ and\ \citenamefont
  {Steele}}]{Castellanos-Gomez2012}%
  \BibitemOpen
  \bibfield  {author} {\bibinfo {author} {\bibfnamefont {A.}~\bibnamefont
  {Castellanos-Gomez}}, \bibinfo {author} {\bibfnamefont {H.~B.}\ \bibnamefont
  {Meerwaldt}}, \bibinfo {author} {\bibfnamefont {W.~J.}\ \bibnamefont
  {Venstra}}, \bibinfo {author} {\bibfnamefont {H.~S.~J.}\ \bibnamefont
  {van~der Zant}}, \ and\ \bibinfo {author} {\bibfnamefont {G.~A.}\
  \bibnamefont {Steele}},\ }\href
  {http://link.aps.org/doi/10.1103/PhysRevB.86.041402} {\bibfield  {journal}
  {\bibinfo  {journal} {Phys. Rev. B}\ }\textbf {\bibinfo {volume} {86}},\
  \bibinfo {pages} {041402} (\bibinfo {year} {2012})}\BibitemShut {NoStop}%
\bibitem [{\citenamefont {Mahboob}\ \emph {et~al.}(2012)\citenamefont
  {Mahboob}, \citenamefont {Nishiguchi}, \citenamefont {Okamoto},\ and\
  \citenamefont {Yamaguchi}}]{Mahboob2012a}%
  \BibitemOpen
  \bibfield  {author} {\bibinfo {author} {\bibfnamefont {I.}~\bibnamefont
  {Mahboob}}, \bibinfo {author} {\bibfnamefont {K.}~\bibnamefont {Nishiguchi}},
  \bibinfo {author} {\bibfnamefont {H.}~\bibnamefont {Okamoto}}, \ and\
  \bibinfo {author} {\bibfnamefont {H.}~\bibnamefont {Yamaguchi}},\ }\href
  {\doibase 10.1038/NPHYS2277} {\bibfield  {journal} {\bibinfo  {journal}
  {Nature Physics}\ }\textbf {\bibinfo {volume} {8}},\ \bibinfo {pages} {387}
  (\bibinfo {year} {2012})}\BibitemShut {NoStop}%
\bibitem [{\citenamefont {Matheny}\ \emph {et~al.}(2013)\citenamefont
  {Matheny}, \citenamefont {Villanueva}, \citenamefont {Karabalin},
  \citenamefont {Sader},\ and\ \citenamefont {Roukes}}]{Matheny2013}%
  \BibitemOpen
  \bibfield  {author} {\bibinfo {author} {\bibfnamefont {M.~H.}\ \bibnamefont
  {Matheny}}, \bibinfo {author} {\bibfnamefont {L.~G.}\ \bibnamefont
  {Villanueva}}, \bibinfo {author} {\bibfnamefont {R.~B.}\ \bibnamefont
  {Karabalin}}, \bibinfo {author} {\bibfnamefont {J.~E.}\ \bibnamefont
  {Sader}}, \ and\ \bibinfo {author} {\bibfnamefont {M.~L.}\ \bibnamefont
  {Roukes}},\ }\href@noop {} {\bibfield  {journal} {\bibinfo  {journal} {Nano
  Lett.}\ }\textbf {\bibinfo {volume} {13}},\ \bibinfo {pages} {1622} (\bibinfo
  {year} {2013})}\BibitemShut {NoStop}%
\bibitem [{\citenamefont {Miao}\ \emph {et~al.}(2014)\citenamefont {Miao},
  \citenamefont {Yeom}, \citenamefont {Wang}, \citenamefont {Standley},\ and\
  \citenamefont {Bockrath}}]{Miao2014}%
  \BibitemOpen
  \bibfield  {author} {\bibinfo {author} {\bibfnamefont {T.~F.}\ \bibnamefont
  {Miao}}, \bibinfo {author} {\bibfnamefont {S.}~\bibnamefont {Yeom}}, \bibinfo
  {author} {\bibfnamefont {P.}~\bibnamefont {Wang}}, \bibinfo {author}
  {\bibfnamefont {B.}~\bibnamefont {Standley}}, \ and\ \bibinfo {author}
  {\bibfnamefont {M.}~\bibnamefont {Bockrath}},\ }\href {\doibase
  10.1021/nl403936a} {\bibfield  {journal} {\bibinfo  {journal} {Nano Lett.}\
  }\textbf {\bibinfo {volume} {14}},\ \bibinfo {pages} {2982} (\bibinfo {year}
  {2014})}\BibitemShut {NoStop}%
\bibitem [{\citenamefont {Sankey}\ \emph {et~al.}(2010)\citenamefont {Sankey},
  \citenamefont {Yang}, \citenamefont {Zwickl}, \citenamefont {Jayich},\ and\
  \citenamefont {Harris}}]{Sankey2010}%
  \BibitemOpen
  \bibfield  {author} {\bibinfo {author} {\bibfnamefont {J.~C.}\ \bibnamefont
  {Sankey}}, \bibinfo {author} {\bibfnamefont {C.}~\bibnamefont {Yang}},
  \bibinfo {author} {\bibfnamefont {B.~M.}\ \bibnamefont {Zwickl}}, \bibinfo
  {author} {\bibfnamefont {A.~M.}\ \bibnamefont {Jayich}}, \ and\ \bibinfo
  {author} {\bibfnamefont {J.~G.~E.}\ \bibnamefont {Harris}},\ }\href
  {http://dx.doi.org/10.1038/nphys1707} {\bibfield  {journal} {\bibinfo
  {journal} {Nat. Phys.}\ }\textbf {\bibinfo {volume} {6}},\ \bibinfo {pages}
  {707} (\bibinfo {year} {2010})}\BibitemShut {NoStop}%
\bibitem [{\citenamefont {Purdy}\ \emph {et~al.}(2010)\citenamefont {Purdy},
  \citenamefont {Brooks}, \citenamefont {Botter}, \citenamefont {Brahms},
  \citenamefont {Ma},\ and\ \citenamefont {Stamper-Kurn}}]{Purdy2010}%
  \BibitemOpen
  \bibfield  {author} {\bibinfo {author} {\bibfnamefont {T.~P.}\ \bibnamefont
  {Purdy}}, \bibinfo {author} {\bibfnamefont {D.~W.~C.}\ \bibnamefont
  {Brooks}}, \bibinfo {author} {\bibfnamefont {T.}~\bibnamefont {Botter}},
  \bibinfo {author} {\bibfnamefont {N.}~\bibnamefont {Brahms}}, \bibinfo
  {author} {\bibfnamefont {Z.-Y.}\ \bibnamefont {Ma}}, \ and\ \bibinfo {author}
  {\bibfnamefont {D.~M.}\ \bibnamefont {Stamper-Kurn}},\ }\href@noop {}
  {\bibfield  {journal} {\bibinfo  {journal} {Phys. Rev. Lett.}\ }\textbf
  {\bibinfo {volume} {105}},\ \bibinfo {pages} {133602} (\bibinfo {year}
  {2010})}\BibitemShut {NoStop}%
\bibitem [{\citenamefont {Aspelmeyer}\ \emph {et~al.}(2014)\citenamefont
  {Aspelmeyer}, \citenamefont {Kippenberg},\ and\ \citenamefont
  {Marquardt}}]{Aspelmeyer2014}%
  \BibitemOpen
  \bibinfo {editor} {\bibfnamefont {M.}~\bibnamefont {Aspelmeyer}}, \bibinfo
  {editor} {\bibfnamefont {T.~J.}\ \bibnamefont {Kippenberg}}, \ and\ \bibinfo
  {editor} {\bibfnamefont {F.}~\bibnamefont {Marquardt}},\ eds.,\ \href@noop {}
  {\emph {\bibinfo {title} {Cavity Optomechanics}}}\ (\bibinfo  {publisher}
  {Springer},\ \bibinfo {year} {Heidelberg, 2014})\BibitemShut {NoStop}%
\bibitem [{\citenamefont {Singh}\ \emph {et~al.}(2014)\citenamefont {Singh},
  \citenamefont {Bosman}, \citenamefont {Schneider}, \citenamefont {Blanter},
  \citenamefont {Castellanos-Gomez},\ and\ \citenamefont
  {{Steele}}}]{Singh2014}%
  \BibitemOpen
  \bibfield  {author} {\bibinfo {author} {\bibfnamefont {V.}~\bibnamefont
  {Singh}}, \bibinfo {author} {\bibfnamefont {S.~J.}\ \bibnamefont {Bosman}},
  \bibinfo {author} {\bibfnamefont {B.~H.}\ \bibnamefont {Schneider}}, \bibinfo
  {author} {\bibfnamefont {Y.~M.}\ \bibnamefont {Blanter}}, \bibinfo {author}
  {\bibfnamefont {A.}~\bibnamefont {Castellanos-Gomez}}, \ and\ \bibinfo
  {author} {\bibfnamefont {G.~A.}\ \bibnamefont {{Steele}}},\ }\href
  {http://dx.doi.org/10.1038/nnano.2014.168} {\bibfield  {journal} {\bibinfo
  {journal} {Nat Nano}\ }\textbf {\bibinfo {volume} {9}},\ \bibinfo {pages}
  {820} (\bibinfo {year} {2014})}\BibitemShut {NoStop}%
\bibitem [{\citenamefont {Weber}\ \emph {et~al.}(2014)\citenamefont {Weber},
  \citenamefont {G\"uttinger}, \citenamefont {Tsioutsios}, \citenamefont
  {Chang},\ and\ \citenamefont {Bachtold}}]{Weber2014}%
  \BibitemOpen
  \bibfield  {author} {\bibinfo {author} {\bibfnamefont {P.}~\bibnamefont
  {Weber}}, \bibinfo {author} {\bibfnamefont {J.}~\bibnamefont {G\"uttinger}},
  \bibinfo {author} {\bibfnamefont {I.}~\bibnamefont {Tsioutsios}}, \bibinfo
  {author} {\bibfnamefont {D.~E.}\ \bibnamefont {Chang}}, \ and\ \bibinfo
  {author} {\bibfnamefont {A.}~\bibnamefont {Bachtold}},\ }\href
  {http://dx.doi.org/10.1021/nl500879k} {\bibfield  {journal} {\bibinfo
  {journal} {Nano Lett.}\ }\textbf {\bibinfo {volume} {14}},\ \bibinfo {pages}
  {2854} (\bibinfo {year} {2014})}\BibitemShut {NoStop}%
\bibitem [{\citenamefont {{Para\"iso}}\ \emph {et~al.}(2015)\citenamefont
  {{Para\"iso}}, \citenamefont {{Kalaee}}, \citenamefont {{Zang}},
  \citenamefont {{Pfeifer}}, \citenamefont {{Marquardt}},\ and\ \citenamefont
  {{Painter}}}]{Paraiso2015}%
  \BibitemOpen
  \bibfield  {author} {\bibinfo {author} {\bibfnamefont {T.~K.}\ \bibnamefont
  {{Para\"iso}}}, \bibinfo {author} {\bibfnamefont {M.}~\bibnamefont
  {{Kalaee}}}, \bibinfo {author} {\bibfnamefont {L.}~\bibnamefont {{Zang}}},
  \bibinfo {author} {\bibfnamefont {H.}~\bibnamefont {{Pfeifer}}}, \bibinfo
  {author} {\bibfnamefont {F.}~\bibnamefont {{Marquardt}}}, \ and\ \bibinfo
  {author} {\bibfnamefont {O.}~\bibnamefont {{Painter}}},\ }\href@noop {}
  {\bibfield  {journal} {\bibinfo  {journal} {ArXiv e-prints}\ } (\bibinfo
  {year} {2015})},\ \Eprint {http://arxiv.org/abs/1505.07291}
  {arXiv:1505.07291} \BibitemShut {NoStop}%
\bibitem [{\citenamefont {{Holland}}\ \emph {et~al.}(2015)\citenamefont
  {{Holland}}, \citenamefont {{Vlastakis}}, \citenamefont {{Heeres}},
  \citenamefont {{Reagor}}, \citenamefont {{Vool}}, \citenamefont {{Leghtas}},
  \citenamefont {{Frunzio}}, \citenamefont {{Kirchmair}}, \citenamefont
  {{Devoret}}, \citenamefont {{Mirrahimi}},\ and\ \citenamefont
  {{Schoelkopf}}}]{Holland2015}%
  \BibitemOpen
  \bibfield  {author} {\bibinfo {author} {\bibfnamefont {E.~T.}\ \bibnamefont
  {{Holland}}}, \bibinfo {author} {\bibfnamefont {B.}~\bibnamefont
  {{Vlastakis}}}, \bibinfo {author} {\bibfnamefont {R.~W.}\ \bibnamefont
  {{Heeres}}}, \bibinfo {author} {\bibfnamefont {M.~J.}\ \bibnamefont
  {{Reagor}}}, \bibinfo {author} {\bibfnamefont {U.}~\bibnamefont {{Vool}}},
  \bibinfo {author} {\bibfnamefont {Z.}~\bibnamefont {{Leghtas}}}, \bibinfo
  {author} {\bibfnamefont {L.}~\bibnamefont {{Frunzio}}}, \bibinfo {author}
  {\bibfnamefont {G.}~\bibnamefont {{Kirchmair}}}, \bibinfo {author}
  {\bibfnamefont {M.~H.}\ \bibnamefont {{Devoret}}}, \bibinfo {author}
  {\bibfnamefont {M.}~\bibnamefont {{Mirrahimi}}}, \ and\ \bibinfo {author}
  {\bibfnamefont {R.~J.}\ \bibnamefont {{Schoelkopf}}},\ }\href@noop {}
  {\bibfield  {journal} {\bibinfo  {journal} {ArXiv e-prints}\ } (\bibinfo
  {year} {2015})},\ \Eprint {http://arxiv.org/abs/1504.03382}
  {arXiv:1504.03382} \BibitemShut {NoStop}%
\bibitem [{\citenamefont {Venstra}\ \emph {et~al.}(2012)\citenamefont
  {Venstra}, \citenamefont {van Leeuwen},\ and\ \citenamefont {van~der
  Zant}}]{Venstra2012}%
  \BibitemOpen
  \bibfield  {author} {\bibinfo {author} {\bibfnamefont {W.~J.}\ \bibnamefont
  {Venstra}}, \bibinfo {author} {\bibfnamefont {R.}~\bibnamefont {van
  Leeuwen}}, \ and\ \bibinfo {author} {\bibfnamefont {H.~S.~J.}\ \bibnamefont
  {van~der Zant}},\ }\href@noop {} {\bibfield  {journal} {\bibinfo  {journal}
  {Appl. Phys. Lett.}\ }\textbf {\bibinfo {volume} {101}},\ \bibinfo {pages}
  {243111} (\bibinfo {year} {2012})}\BibitemShut {NoStop}%
\bibitem [{\citenamefont {Vinante}(2014)}]{Vinante2014}%
  \BibitemOpen
  \bibfield  {author} {\bibinfo {author} {\bibfnamefont {A.}~\bibnamefont
  {Vinante}},\ }\href@noop {} {\bibfield  {journal} {\bibinfo  {journal} {Phys.
  Rev. B}\ }\textbf {\bibinfo {volume} {90}},\ \bibinfo {pages} {024308}
  (\bibinfo {year} {2014})}\BibitemShut {NoStop}%
\bibitem [{\citenamefont {Santamore}\ \emph {et~al.}(2004)\citenamefont
  {Santamore}, \citenamefont {Doherty},\ and\ \citenamefont
  {Cross}}]{Santamore2004a}%
  \BibitemOpen
  \bibfield  {author} {\bibinfo {author} {\bibfnamefont {D.~H.}\ \bibnamefont
  {Santamore}}, \bibinfo {author} {\bibfnamefont {A.~C.}\ \bibnamefont
  {Doherty}}, \ and\ \bibinfo {author} {\bibfnamefont {M.~C.}\ \bibnamefont
  {Cross}},\ }\href@noop {} {\bibfield  {journal} {\bibinfo  {journal} {Phys.
  Rev. B}\ }\textbf {\bibinfo {volume} {70}},\ \bibinfo {pages} {144301}
  (\bibinfo {year} {2004})}\BibitemShut {NoStop}%
\bibitem [{\citenamefont {Ludwig}\ \emph {et~al.}(2012)\citenamefont {Ludwig},
  \citenamefont {Safavi-Naeini}, \citenamefont {Painter},\ and\ \citenamefont
  {Marquardt}}]{Ludwig2012}%
  \BibitemOpen
  \bibfield  {author} {\bibinfo {author} {\bibfnamefont {M.}~\bibnamefont
  {Ludwig}}, \bibinfo {author} {\bibfnamefont {A.~H.}\ \bibnamefont
  {Safavi-Naeini}}, \bibinfo {author} {\bibfnamefont {O.}~\bibnamefont
  {Painter}}, \ and\ \bibinfo {author} {\bibfnamefont {F.}~\bibnamefont
  {Marquardt}},\ }\href@noop {} {\bibfield  {journal} {\bibinfo  {journal}
  {Phys. Rev. Lett.}\ }\textbf {\bibinfo {volume} {109}},\ \bibinfo {pages}
  {063601} (\bibinfo {year} {2012})}\BibitemShut {NoStop}%
\bibitem [{\citenamefont {Dykman}\ and\ \citenamefont
  {Krivoglaz}(1984)}]{DK_review84}%
  \BibitemOpen
  \bibfield  {author} {\bibinfo {author} {\bibfnamefont {M.~I.}\ \bibnamefont
  {Dykman}}\ and\ \bibinfo {author} {\bibfnamefont {M.~A.}\ \bibnamefont
  {Krivoglaz}},\ }in\ \href@noop {} {\emph {\bibinfo {booktitle} {Sov. Phys.
  Reviews}}},\ Vol.~\bibinfo {volume} {5},\ \bibinfo {editor} {edited by\
  \bibinfo {editor} {\bibfnamefont {I.~M.}\ \bibnamefont {Khalatnikov}}}\
  (\bibinfo  {publisher} {Harwood Academic, New York},\ \bibinfo {year}
  {1984})\ pp.\ \bibinfo {pages} {265--441,
  http://www.pa.msu.edu/~dykman/pub06/DKreview84.pdf}\BibitemShut {NoStop}%
\bibitem [{\citenamefont {{Sansa}}\ \emph {et~al.}(2015)\citenamefont
  {{Sansa}}, \citenamefont {{Sage}}, \citenamefont {{Bullard}}, \citenamefont
  {{Gely}}, \citenamefont {{Alava}}, \citenamefont {{Colinet}}, \citenamefont
  {{Naik}}, \citenamefont {{Villanueva}}, \citenamefont {{Duraffourg}},
  \citenamefont {{Roukes}}, \citenamefont {{Jourdan}},\ and\ \citenamefont
  {{Hentz}}}]{Sansa2015}%
  \BibitemOpen
  \bibfield  {author} {\bibinfo {author} {\bibfnamefont {M.}~\bibnamefont
  {{Sansa}}}, \bibinfo {author} {\bibfnamefont {E.}~\bibnamefont {{Sage}}},
  \bibinfo {author} {\bibfnamefont {E.~C.}\ \bibnamefont {{Bullard}}}, \bibinfo
  {author} {\bibfnamefont {M.}~\bibnamefont {{Gely}}}, \bibinfo {author}
  {\bibfnamefont {T.}~\bibnamefont {{Alava}}}, \bibinfo {author} {\bibfnamefont
  {E.}~\bibnamefont {{Colinet}}}, \bibinfo {author} {\bibfnamefont {A.~K.}\
  \bibnamefont {{Naik}}}, \bibinfo {author} {\bibfnamefont {G.~L.}\
  \bibnamefont {{Villanueva}}}, \bibinfo {author} {\bibfnamefont
  {L.}~\bibnamefont {{Duraffourg}}}, \bibinfo {author} {\bibfnamefont {M.~L.}\
  \bibnamefont {{Roukes}}}, \bibinfo {author} {\bibfnamefont {G.}~\bibnamefont
  {{Jourdan}}}, \ and\ \bibinfo {author} {\bibfnamefont {S.}~\bibnamefont
  {{Hentz}}},\ }\href@noop {} {\bibfield  {journal} {\bibinfo  {journal} {ArXiv
  e-prints}\ } (\bibinfo {year} {2015})},\ \Eprint
  {http://arxiv.org/abs/1506.08135} {arXiv:1506.08135} \BibitemShut {NoStop}%
\bibitem [{\citenamefont {{Zhang}}\ \emph {et~al.}(2014)\citenamefont
  {{Zhang}}, \citenamefont {{Moser}}, \citenamefont {G\"uttinger},
  \citenamefont {{Bachtold}},\ and\ \citenamefont {{Dykman}}}]{Zhang2014}%
  \BibitemOpen
  \bibfield  {author} {\bibinfo {author} {\bibfnamefont {Y.}~\bibnamefont
  {{Zhang}}}, \bibinfo {author} {\bibfnamefont {J.}~\bibnamefont {{Moser}}},
  \bibinfo {author} {\bibfnamefont {J.}~\bibnamefont {G\"uttinger}}, \bibinfo
  {author} {\bibfnamefont {A.}~\bibnamefont {{Bachtold}}}, \ and\ \bibinfo
  {author} {\bibfnamefont {M.~I.}\ \bibnamefont {{Dykman}}},\ }\href@noop {}
  {\bibfield  {journal} {\bibinfo  {journal} {Phys. Rev. Lett.}\ }\textbf
  {\bibinfo {volume} {113}},\ \bibinfo {pages} {255502} (\bibinfo {year}
  {2014})}\BibitemShut {NoStop}%
\bibitem [{\citenamefont {Lorentz}(1916)}]{Lorentz1916}%
  \BibitemOpen
  \bibfield  {author} {\bibinfo {author} {\bibfnamefont {H.~A.}\ \bibnamefont
  {Lorentz}},\ }\href@noop {} {\emph {\bibinfo {title} {The theory of electrons
  and its applications to the phenomena of light and radiant heat}}}\ (\bibinfo
   {publisher} {Teubner, B. G.},\ \bibinfo {year} {Leipzig, 1916})\BibitemShut
  {NoStop}%
\bibitem [{\citenamefont {Einstein}\ and\ \citenamefont
  {Hopf}(1910)}]{Einstein1910b}%
  \BibitemOpen
  \bibfield  {author} {\bibinfo {author} {\bibfnamefont {A.}~\bibnamefont
  {Einstein}}\ and\ \bibinfo {author} {\bibfnamefont {L.}~\bibnamefont
  {Hopf}},\ }\href@noop {} {\bibfield  {journal} {\bibinfo  {journal} {Ann.d.
  Phys.}\ }\textbf {\bibinfo {volume} {33}},\ \bibinfo {pages} {1105} (\bibinfo
  {year} {1910})}\BibitemShut {NoStop}%
\bibitem [{\citenamefont {Heitler}(2010)}]{Heitler2010}%
  \BibitemOpen
  \bibfield  {author} {\bibinfo {author} {\bibfnamefont {W.}~\bibnamefont
  {Heitler}},\ }\href@noop {} {\emph {\bibinfo {title} {The Quantum Theory of
  Radiation, 3rd ed.}}}\ (\bibinfo  {publisher} {Dover Publications, Inc., New
  York},\ \bibinfo {year} {2010})\BibitemShut {NoStop}%
\bibitem [{\citenamefont {Dykman}\ and\ \citenamefont
  {Krivoglaz}(1971)}]{Dykman1971}%
  \BibitemOpen
  \bibfield  {author} {\bibinfo {author} {\bibfnamefont {M.~I.}\ \bibnamefont
  {Dykman}}\ and\ \bibinfo {author} {\bibfnamefont {M.~A.}\ \bibnamefont
  {Krivoglaz}},\ }\href@noop {} {\bibfield  {journal} {\bibinfo  {journal}
  {Phys. Stat. Sol. B}\ }\textbf {\bibinfo {volume} {48}},\ \bibinfo {pages}
  {497} (\bibinfo {year} {1971})}\BibitemShut {NoStop}%
\bibitem [{\citenamefont {Anderson}(1954)}]{Anderson1954}%
  \BibitemOpen
  \bibfield  {author} {\bibinfo {author} {\bibfnamefont {P.~W.}\ \bibnamefont
  {Anderson}},\ }\href@noop {} {\bibfield  {journal} {\bibinfo  {journal} {J.
  Phys. Soc. Japan}\ }\textbf {\bibinfo {volume} {9}},\ \bibinfo {pages} {316}
  (\bibinfo {year} {1954})}\BibitemShut {NoStop}%
\bibitem [{\citenamefont {Kubo}(1954)}]{Kubo1954}%
  \BibitemOpen
  \bibfield  {author} {\bibinfo {author} {\bibfnamefont {R.}~\bibnamefont
  {Kubo}},\ }\href@noop {} {\bibfield  {journal} {\bibinfo  {journal} {J. Phys.
  Soc. Japan}\ }\textbf {\bibinfo {volume} {9}},\ \bibinfo {pages} {935}
  (\bibinfo {year} {1954})}\BibitemShut {NoStop}%
\bibitem [{\citenamefont {Senitzky}(1960)}]{Senitzky1960}%
  \BibitemOpen
  \bibfield  {author} {\bibinfo {author} {\bibfnamefont {I.~R.}\ \bibnamefont
  {Senitzky}},\ }\href@noop {} {\bibfield  {journal} {\bibinfo  {journal}
  {Phys. Rev.}\ }\textbf {\bibinfo {volume} {119}},\ \bibinfo {pages} {670}
  (\bibinfo {year} {1960})}\BibitemShut {NoStop}%
\bibitem [{\citenamefont {Schwinger}(1961)}]{Schwinger1961}%
  \BibitemOpen
  \bibfield  {author} {\bibinfo {author} {\bibfnamefont {J.}~\bibnamefont
  {Schwinger}},\ }\href@noop {} {\bibfield  {journal} {\bibinfo  {journal} {J.
  Math. Phys.}\ }\textbf {\bibinfo {volume} {2}},\ \bibinfo {pages} {407}
  (\bibinfo {year} {1961})}\BibitemShut {NoStop}%
\bibitem [{\citenamefont {Louisell}(1990)}]{Louisell1990}%
  \BibitemOpen
  \bibfield  {author} {\bibinfo {author} {\bibfnamefont {W.~H.}\ \bibnamefont
  {Louisell}},\ }\href@noop {} {\emph {\bibinfo {title} {Quantum Statistical
  Properties of Radiation}}}\ (\bibinfo  {publisher} {Wiley-VCH},\ \bibinfo
  {year} {Berlin, 1990})\BibitemShut {NoStop}%
\bibitem [{\citenamefont {Gelfand}\ and\ \citenamefont
  {Yaglom}(1960)}]{Gelfand1960}%
  \BibitemOpen
  \bibfield  {author} {\bibinfo {author} {\bibfnamefont {I.~M.}\ \bibnamefont
  {Gelfand}}\ and\ \bibinfo {author} {\bibfnamefont {A.~M.}\ \bibnamefont
  {Yaglom}},\ }\href@noop {} {\bibfield  {journal} {\bibinfo  {journal} {J.
  Math. Phys.}\ }\textbf {\bibinfo {volume} {1}},\ \bibinfo {pages} {48}
  (\bibinfo {year} {1960})}\BibitemShut {NoStop}%
\bibitem [{\citenamefont {Feynman}\ and\ \citenamefont
  {Hibbs}(1965)}]{FeynmanQM}%
  \BibitemOpen
  \bibfield  {author} {\bibinfo {author} {\bibfnamefont {R.~P.}\ \bibnamefont
  {Feynman}}\ and\ \bibinfo {author} {\bibfnamefont {A.~R.}\ \bibnamefont
  {Hibbs}},\ }\href@noop {} {\emph {\bibinfo {title} {Quantum Mechanics and
  Path Integrals}}}\ (\bibinfo  {publisher} {McGraw-Hill},\ \bibinfo {address}
  {New-York},\ \bibinfo {year} {1965})\BibitemShut {NoStop}%
\bibitem [{\citenamefont {Phythian}(1977)}]{Phythian1977}%
  \BibitemOpen
  \bibfield  {author} {\bibinfo {author} {\bibfnamefont {R.}~\bibnamefont
  {Phythian}},\ }\href@noop {} {\bibfield  {journal} {\bibinfo  {journal} {J.
  Phys. A}\ }\textbf {\bibinfo {volume} {10}},\ \bibinfo {pages} {777}
  (\bibinfo {year} {1977})}\BibitemShut {NoStop}%
\bibitem [{\citenamefont {Mannella}(2002)}]{Mannella2002a}%
  \BibitemOpen
  \bibfield  {author} {\bibinfo {author} {\bibfnamefont {R.}~\bibnamefont
  {Mannella}},\ }\href@noop {} {\bibfield  {journal} {\bibinfo  {journal} {Int.
  J. Mod. Phys. C}\ }\textbf {\bibinfo {volume} {13}},\ \bibinfo {pages} {1177}
  (\bibinfo {year} {2002})}\BibitemShut {NoStop}%
\bibitem [{\citenamefont {Dykman}(2012)}]{Dykman2012b}%
  \BibitemOpen
  \bibinfo {editor} {\bibfnamefont {M.~I.}\ \bibnamefont {Dykman}},\ ed.,\
  \href@noop {} {\emph {\bibinfo {title} {Fluctuating Nonlinear Oscillators:
  from Nanomechanics to Quantum Superconducting Circuits}}}\ (\bibinfo
  {publisher} {OUP, Oxford},\ \bibinfo {year} {2012})\BibitemShut {NoStop}%
\bibitem [{\citenamefont {Zhang}\ \emph {et~al.}(2014)\citenamefont {Zhang},
  \citenamefont {Tadokoro},\ and\ \citenamefont {{Dykman}}}]{Zhang2014a}%
  \BibitemOpen
  \bibfield  {author} {\bibinfo {author} {\bibfnamefont {Y.}~\bibnamefont
  {Zhang}}, \bibinfo {author} {\bibfnamefont {Y.}~\bibnamefont {Tadokoro}}, \
  and\ \bibinfo {author} {\bibfnamefont {M.~I.}\ \bibnamefont {{Dykman}}},\
  }\href@noop {} {\bibfield  {journal} {\bibinfo  {journal} {NJP}\ }\textbf
  {\bibinfo {volume} {16}},\ \bibinfo {pages} {113064} (\bibinfo {year}
  {2014})}\BibitemShut {NoStop}%
\bibitem [{\citenamefont {Landau}\ and\ \citenamefont
  {Lifshitz}(2004)}]{LL_Mechanics2004}%
  \BibitemOpen
  \bibfield  {author} {\bibinfo {author} {\bibfnamefont {L.~D.}\ \bibnamefont
  {Landau}}\ and\ \bibinfo {author} {\bibfnamefont {E.~M.}\ \bibnamefont
  {Lifshitz}},\ }\href@noop {} {\emph {\bibinfo {title} {Mechanics}}},\
  \bibinfo {edition} {3rd}\ ed.\ (\bibinfo  {publisher} {Elsevier, Amsterdam},\
  \bibinfo {year} {2004})\BibitemShut {NoStop}%
\bibitem [{\citenamefont {Dykman}\ and\ \citenamefont
  {Krivoglaz}(1979)}]{Dykman1979a}%
  \BibitemOpen
  \bibfield  {author} {\bibinfo {author} {\bibfnamefont {M.~I.}\ \bibnamefont
  {Dykman}}\ and\ \bibinfo {author} {\bibfnamefont {M.~A.}\ \bibnamefont
  {Krivoglaz}},\ }\href@noop {} {\bibfield  {journal} {\bibinfo  {journal} {Zh.
  Eksp. Teor. Fiz.}\ }\textbf {\bibinfo {volume} {77}},\ \bibinfo {pages} {60}
  (\bibinfo {year} {1979})}\BibitemShut {NoStop}%
\bibitem [{\citenamefont {Drummond}\ and\ \citenamefont
  {Walls}(1980)}]{Drummond1980c}%
  \BibitemOpen
  \bibfield  {author} {\bibinfo {author} {\bibfnamefont {P.~D.}\ \bibnamefont
  {Drummond}}\ and\ \bibinfo {author} {\bibfnamefont {D.~F.}\ \bibnamefont
  {Walls}},\ }\href@noop {} {\bibfield  {journal} {\bibinfo  {journal} {J.
  Phys. A}\ }\textbf {\bibinfo {volume} {13}},\ \bibinfo {pages} {725}
  (\bibinfo {year} {1980})}\BibitemShut {NoStop}%
\end{thebibliography}

%

\end{document}